\documentclass[twocolumn, longauthor]{aastex61}

\newcommand{\caii}{Ca~{\sc ii}}
\newcommand{\caiihk}{Ca~{\sc ii} H \& K}
\newcommand{\nai}{Na~{\sc i}}
\newcommand{\nad}{Na~{\sc i}~D}
\newcommand{\ki}{K~{\sc i}}

\newcommand{\smw}{\ensuremath{S_{\rm MW}}}
\newcommand{\rphk}{\ensuremath{R'_{\rm HK}}}
\newcommand{\lrphk}{\ensuremath{\log{\rphk}}} 

\newcommand{\feh}{\ensuremath{[\mbox{Fe}/\mbox{H}]}}
\newcommand{\teff}{\ensuremath{T_{\mbox{\scriptsize eff}}}}
\newcommand{\logg}{\ensuremath{\log g}}

\newcommand{\vsini}{\ensuremath{v \sin i}}
\newcommand{\kms}{\ensuremath{\mbox{km s}^{-1}}}
\newcommand{\ma}{m\AA}
\newcommand{\soren}{S\o ren Meibom}

\received{2016 May 17}
\revised{2017 May 3}
\accepted{2017 May 12}

\submitjournal{AJ}

\shorttitle{No ``Maunder minimum'' candidates in M67}

\shortauthors{Curtis}

\begin{document}

\title{No ``Maunder minimum'' candidates in M67: 
mitigating interstellar contamination of chromospheric emission lines}

\email{jasoncurtis.astro@gmail.com}

\author{Jason Lee Curtis}
\altaffiliation{NSF Astronomy \& Astrophysics Postdoctoral Fellow}
\affiliation{Department of Astronomy, Columbia University, 550 West 120th Street, New York, NY 10027, USA}
\affiliation{Harvard--Smithsonian Center for Astrophysics, 60 Garden Street, Cambridge, MA 02138, USA}
\affiliation{Center for Exoplanets and Habitable Worlds, Department of Astronomy \& Astrophysics, 
    The Pennsylvania State University, \\ 
    525 Davey Laboratory, University Park, PA 16802, USA}



\begin{abstract}
The solar analogs of M67 let us glimpse the probable behavior of the Sun 
on time scales surpassing the duration of human civilization. 
M67 can serve as a solar proxy because its stars share a similar age and 
composition with the Sun. 
Previous surveys of M67 observed that 15\% of its Sun-like stars exhibited chromospheric activity levels 
below solar minimum, 
which suggest that these stars might be in activity-minimum states analogous 
to the Maunder Minimum. 
The activity diagnostic used, the HK index 
(relative intensities of the \caiihk\ lines integrated over 1~\AA\ bandpasses), 
was measured from low-resolution spectra ($R\approx 5000$), 
as is traditional and suitable for nearby, bright stars. 
However, for stars beyond the Local Bubble, 
the interstellar medium (ISM) imprints absorption lines in spectra at \caiihk , 
which negatively bias activity measurements when these lines 
fall within the HK index bandpass. 
I model the ISM clouds in the M67 foreground 
with high-resolution spectra of blue stragglers and solar analogs. 
I demonstrate that ISM absorption varies across the cluster
and must be accounted for on a star-by-star basis.
I then apply the ISM model to a solar spectrum
and broaden it to the lower spectral resolution employed by prior surveys.
Comparing HK indices measured from ISM-free and ISM-contaminated spectra, 
I find that all stars observed below solar minimum can be explained by this ISM bias. 
I conclude that there is no compelling evidence for Maunder minimum 
candidates in M67 at this time. 
\end{abstract}






\section{Introduction} \label{s:intro}

The Sun obviously influences Earth's climate and habitability, 
and yet our ability to explore and fully comprehend this Sun--Earth connection is inhibited  
by the Sun's dynamic nature over a wide range of time scales. 
Solar physicists have only been able to study the UV and X-ray emission 
originating from the Sun's magnetically heated outer atmosphere 
over the last 50 years since the advent of space-based observatories.
The sunspot record dates back over 400 years to Galileo and  
has revealed rare events like the Maunder Minimum---the 70 year period (1645--1715) 
when sunspots rarely formed 
\citep[for a forceful argument for the reality and importance of the Maunder Minimum, 
see][]{Eddy1976}.
Measurements of cosmogenic radionuclides from ice core ($^{10}$Be) and tree ring ($^{14}$C) samples 
trace the history of solar activity based on how the heliosphere
modulates cosmic rays.
These data, corrected for the influence of Earth's magnetosphere, 
take our view of the solar dynamo back over 10,000 years 
\citep[for a review, see][]{Usoskin2013}, 
and yet this baseline barely encompasses the birth of human civilization, 
whereas anatomically modern humans evolved some 195,000 years ago
\citep{omoage}.

To understand how the activity of the contemporary Sun has influenced biological evolution 
and habitability over the last 500 Myr since the Cambrian explosion, 
and to properly constrain the frequency and appreciate the potential impact of grand activity maxima and minima, 
we must look to the stars.

The behaviors of stars that are similar in mass and composition to the Sun 
(i.e., solar analogs and twins) 
allow us to use modern astronomical observatories to study 
the Sun's likely activity and variability on much longer timescales 
along with its evolution over cosmic time,
and investigate the frequency of aperiodic events like the aforementioned Maunder Minimum.
Bright, nearby stars have been surveyed for photospheric and chromospheric
variability for decades, for example, by the Mount Wilson Observatory HK Project 
\citep{Wilson1968, Vaughan1978, Baliunas1998}.
The upper chromospheres and transition regions of these bright stars are easily accessible with FUV spectroscopy 
with \textit{HST}, while \textit{Chandra} provides a view of their 
soft X-ray emitting coronae  \citep[e.g., the Sun in Time project,][]{SunInTime1}.
However, our ability to relate the rotation, XUV emission, 
and optical diagnostics of magnetic activity (e.g., \caiihk , H$\alpha$, the Ca IR triplet) 
to a particular era relevant to Earth's early biological evolution or the modern day
is limited by the large age uncertainties common for individual field stars. 

Not only are ages for individual Sun-like stars notoriously difficult to infer
\citep{soderblom2010}, 
without reliable parallaxes and metallicities it can even be difficult to establish their 
evolutionary states (i.e., dwarf vs. subgiant). 
This challenge plagued the early analyses of the Mount Wilson Observatory HK Project sample, 
where \citet{Baliunas1990} concluded that approximately one-third of Sun-like stars
are in grand minima states analogous to the Maunder Minimum (i.e., ``Maunder minima'' stars, 
denoted MM, and where minimum is in lowercase to distinguish these analogous states from 
the actual solar Maunder Minimum period). 
Using accurate parallaxes from  \textit{Hipparcos}, 
\citet{wright_maunder} demonstrated that most or all of these MM candidates 
are in fact old and have evolved off the main sequence, and are therefore not analogous to the Sun.

Benchmark star clusters provide large samples of stars with 
accurate and precise ages (though limited by the ultimate accuracy of stellar evolution models)
and therefore complement the 
sparse sample of individual nearby stars 
with well-established ages (e.g., from asteroseismology or differential spectroscopy).
Clusters are superb laboratories for exploring the evolution
of angular momentum and magnetic activity in cool stars. Their
advantage lies in providing a population of stars of the
same age and metallicity spanning a wide range of mass, thus allowing
some control over the variables involved, 
while providing opportunities complementary to the studies of individual, nearby stars and solar twins.
\begin{enumerate}
 \item Typical activity levels can be instantaneously mapped as a function of mass 
      by averaging over the random cycle phases for individual stars in a given mass bin.
 \item Estimates for cycle ranges can be derived from the scatter about these averages.
 \item The impact of tidally interacting binaries (with stellar or sub-stellar companions) 
       on rotation--activity evolution can be assessed as a function of orbital period.
 \item Establishing cluster activity sequences facilitates the identification of 
 outliers potentially existing in grand maxima and minima states. 
 Completing a thorough survey of benchmark clusters with a range of ages would
 help constrain the frequency of these events over cosmic time.
\end{enumerate}

\subsection{M67 As a Solar Benchmark}
For cluster-based studies of the modern Sun, M67 stands without rival. 
M67 is the prototypical old open cluster \citep{classicM67}. 
Its age and composition are nearly indistinguishable from the Sun \citep{M67SolarTwin}, 
which led some to question whether M67 was the Sun's natal cluster.
However, \citet{sunM67} concluded that this scenario was unlikely
based on a dynamical analysis of the Sun and M67's motion through a realistic Milky Way gravitational potential. 
Crucially, the membership and multiplicity of M67 have been carefully determined from 
over 40 years of radial velocity monitoring \citep{Geller2015, LathamM67}.

M67 is thus an excellent target for understanding the magnetic activity 
of the Sun and Sun-like stars at 4 Gyr. 
It was surveyed for 75 consecutive days in 2015 with NASA's re-purposed \textit{Kepler} mission,
\textit{K2}, during Campaign 5,
and rotation periods have already been measured for a few dozen FGK dwarfs
\citep{Barnes2016, Gonzalez2016}.
Regarding chromospheric activity, 
\citet{Baliunas1995-relation} and \citet{Giampapa2006} utilized the 
Hydra multiobject spectrograph \citep{Barden1995} 
to measure \caiihk\ activity indices for 77 solar-type stars on multiple occasions over a six year period.
At a distance of 880 pc, the solar analogs of M67 are relatively faint ($14<V<15.3$), 
which makes obtaining high-resolution optical spectra at high signal-to-noise ratios ($S/N$) for these stars costly. 
This difficulty is compounded when one attempts to measure 
the weak chromospheric emission in the deep \caiihk\ line cores.
\citeauthor{Giampapa2006} opted for a low-resolution setup for Hydra ($R\approx 5000$) so that the
chromospheric \caiihk\ emission could be measured at high enough precision to discern variability 
over the multi-year survey. 

\citet{Giampapa2006} found that $\sim$17\% of M67's solar analogs showed average activity levels 
in deficit of solar minimum. 
Furthermore, about 7--12\% exhibited activity in excess of solar maximum. 
These observational results empirically suggest that the Sun can likewise enter high- and low-activity 
states at similar frequencies. 
In fact, according to the sunspot number over the last 11,000 years that was reconstructed from the 
radionuclide record, the Sun has operated in grand minima for 17\% and in grand maxima for 10\% 
of the modern era \citep{Usoskin2013}.\footnote{Of the 27 grand minima identified in the 
previous 11,000 years, the duration ranges between 20 and 160 years with a median duration of 60 years.} 
These numbers appear in remarkable agreement with the results from M67.
This conclusion has profound implications for understanding the Sun's past and potential impact 
on Earth's climate and habitability. 

However, as I discuss in this work, the low-resolution Hydra spectroscopy suffers 
from two important astrophysical systematic biases.
First, at a distance of 880~pc, there is a high likelihood for intervening 
interstellar gas to form absorption lines near the stellar \caiihk\ line cores, 
which would contaminate the chromospheric activity measurements. 
This problem was recognized by \citet{Pace2004} \citep[see also;][]{Pace2009}, 
and I develop this further by illustrating the presence of 
a spatially varying interstellar medium (ISM) and 
demonstrating how this ISM bias explains the appearance of M67 stars with activity 
levels below Solar minimum.\footnote{\citet{Fossati2017} explore this same 
ISM bias as an explanation for the appearance 
of low-activity levels in particular hot Jupiter hosts (e.g., WASP-13).}
Second, much progress has been made in identifying binaries in the years since \citet{Giampapa2006}.
Utilizing \citeauthor{Geller2015}'s \citeyear{Geller2015} new membership and binarity catalog, 
I find that all but one of the stars with activity levels in excess of solar maximum are binaries, 
which suggests that their angular momentum evolution has been influenced by tidal interaction with 
a companion, 
that the line core emission has been enhanced by blending from two sources, 
or that the primary's motion relative to the ISM has swept varying 
amounts of the ISM lines into the HK bandpass over time.

\section{The ISM toward M67} \label{s:ism}
Interstellar gas is observable 
from atomic absorption by
\caii\ (H \& K at 3968.469 and 3933.663~\AA),
\nai\ (D2 at 5889.951~\AA, D1 at 5895.924~\AA), and
\ki\ (7698.964~\AA).
The equivalent widths of these prominent absorption lines 
generally correlate with the amount of foreground dust and reddening, 
as measured from OB stellar \citep{Munari1997}
and quasar spectra \citep{Poznanski2012}.\footnote{While 
the strength of interstellar absorption lines have been found to 
generally correlate with visual extinction and reddening, it is not known whether and to what 
degree these line strengths are sensitive to small differences in $E(B-V)$ and $A_V$, 
which trace the line-of-sight dust content, 
which is different than the gas that produces the ISM lines. 
Evidence for differential reddening has been 
presented for NGC 6819 
\citep[the 2.5 Gyr \textit{Kepler} cluster,][]{Platais6819,Twarog6819}.
The WIYN Open Cluster Study (WOCS) obtained spectra for these clusters 
that were used to measure barium abundances 
to test the AGB mass transfer formation scenario 
for that cluster's blue straggler population. 
The Ba spectral order encompasses \nad, 
which means those spectra can be analyzed to search 
for variable ISM absorption
and check for spatial correlation with the  
differential reddening map suggested by \citet{Platais6819} and \citet{Twarog6819}.}

I use high-resolution optical spectroscopy
to identify interstellar clouds in the M67 foreground 
from \nad\ absorption.
While this paper is principally concerned with the interstellar contamination
of \caiihk\ activity indices, 
the stellar \nad\ lines are more easily modeled than \caiihk , 
which facilitates the identification of individual interstellar clouds.
After identifying these clouds, 
I model their \caiihk\ absorption lines using blue straggler stars (BSS)
that effectively backlight the ISM. 
BSS spectra often have fewer, broadened features compared to Sun-like stars 
due to their generally 
hotter effective temperatures and rapid rotation. 
Therefore, narrow lines due to interstellar clouds 
and telluric absorption
can be isolated, 
then corrected for in the more complex spectra of Sun-like stars.

\subsection{Solar activity spectra and indices} \label{s:solardata}
This study requires high-resolution optical spectra of the 
Sun and a sample of M67 members.
For the Sun, 
I used the \citet{Wallace2011} disk-averaged Solar Flux atlas,
and the telluric spectrum derived from disk-center observations taken at 
different airmasses.\footnote{Described at \url{http://diglib.nso.edu/flux} and accessible at \url{ftp://vso.nso.edu/pub/Wallace_2011_solar_flux_atlas}}

The Sun has also been observed regularly at a spectral 
resolution of $R \approx 300,000$ since 2006 December (JD 2454072)
by the National Solar Observatory's 
Synoptic Optical Long-term Investigations of the Sun (SOLIS) facility
with the Integrated Sunlight Spectrometer (ISS) on Kitt Peak \citep{solis}. 
This survey observes important spectral regions, including 
\caiihk, 
and the spectra are downloadable from the 
SOLIS website.\footnote{\url{http://solis.nso.edu/iss}} 
I apply my ISM model for M67 to these ISS spectra in order to derive 
a correction to the HK indices from \citet{Giampapa2006}.

The SOLIS website also provides tabulated 
chromospheric H \& K indices with $\sim$0.001\%. measurement errors 
\citep[procedure described by][]{issK}, 
including the 1~\AA\ indices used by \citet{Giampapa2006}. 
However, these measurements are not on the same system as 
the M67 data. 
\citet{Egeland2016} describe the various available solar activity 
data sets and provide equations to transform between the different 
systems. \citet{Giampapa2006} calibrated their M67 data following 
\citet{HallLockwood1995} in order to place their HK indices on the 
same system as the solar activity data from the
NSO McMath--Pierce Solar Telescope on Kitt Peak 
\citep[KP;][]{Livingston2007}\footnote{\url{ftp://vso.nso.edu/cycle_spectra/reduced_data/fd.log2}}
when scaled by 4.7\% to account for differences in spectral resolution.
\citet{Egeland2016} derived $K_{\rm KP} = 1.143\, K_{\rm SP} - 0.0148$, 
and \citet{Bertello2017} give 
$K_{\rm ISS} = 0.8781\, K_{\rm SP} + 0.0062$ to transform 
between the NSO Sacramento Peak \citep[SP;][]{SacPeak1, SacPeak2} 
and ISS records. 

\caii~K filtergrams of the Sun dating back to 1907 
are available from the Kodaikanal Solar Observatory in India (KKL). 
Using this record, \citet{Bertello2016} and \citet{Bertello2017} 
produce a composite solar K index history for the past 110 years. 
In order to compare the range in solar activity with the M67 distribution,
I use the monthly averaged ISS/SP/KKL \caii~K~1~\AA\ emission index 
time series,\footnote{\url{http://solis.nso.edu/0/iss/composite.dat}}
which is calibrated to the $K_{\rm ISS}$ system, 
then I transform the series to $K_{\rm KP}$ by combining the equations 
relating $K_{\rm ISS}$ to $K_{\rm SP}$ and $K_{\rm SP}$ to $K_{\rm KP}$, 
then convert to $HK_{\rm KP}$ by applying a linear relationship between 
$H_{\rm KP}$ and $K_{\rm KP}$ that I derived using the NSO Kitt Peak data set: 
$H_{\rm KP} = 0.7863\, K_{\rm KP} + 0.0237$.\footnote{For this linear regression, 
I used the KP data set from 1976 to 1993, 
where there is an apparent discontinuity in the ratio between the H and K indices.} 
Finally, I scale this series by 1.047 to account for the difference in spectral resolution, 
as noted and applied by \citet{Giampapa2006}. 
This procedure yields the monthly average HK value for the Sun from 1907 to the present day 
that is directly comparable to the HK indices for M67 published by \citet{Giampapa2006}.
On this system, I find that the Sun ranged from 182 to 219~m\AA , 
with a median and standard deviation of 193 and 8~m\AA .
Using the full ISS/SP record from 1976 to present extends the range 
from 179 to 226~\ma .\footnote{\url{http://solis.nso.edu/0/iss/sp_iss.dat}} 
\citet{Giampapa2006} adopted 225~\ma\ as solar maximum using the Kitt Peak series; 
I will classify M67 members as overactive or underactive as those departing from 
the solar range of 179 to 226~\ma , which corresponds to 
$S$ = 0.1586 to 0.1916 and $\lrphk = -5.005$ to $-4.834$~dex.\footnote{For 
$S$ and \lrphk , I followed the above equations to convert to $K_{\rm SP}$, 
the \citet{Egeland2016} relation to convert to $S$ ($S = 1.5\, K_{\rm SP} + 0.031$), 
and the \citet{Noyes1984} procedure to convert to \lrphk, adopting $(B - V) = 0.65$ for the solar color; 
\citep[see also the top panel of Figure~4 from][]{Egeland2016}.}


\subsection{High-resolution optical spectroscopy for M67} \label{s:M67data}

For the M67 sample, 
I use the spectra acquired and analyzed by \citet{Pace2004}, 
including the blue straggler Sanders 1082 
(S1082, F131, observed on 2002 March 19 at 
$S/N \approx 50$ at \caiihk, and $S/N \approx 40$ at \nad), 
and the solar analogs S746, S1048, and S1255. 
Each of these solar analogs was observed twice at $S/N \approx 14-19$. 
While these per-pixel values might seem low, 
integrating over the 1~\AA\ bandpasses for the H \& K indices
averaged out the noise to yield measurements of sufficient accuracy 
for the \citet{Pace2004} analysis. 
However, modeling the ISM lines in the deep \caiihk\ line cores requires a higher quality 
spectrum. I generated this by co-adding the six spectra for the three solar 
analogs.\footnote{There are actually observations of seven 
solar-type stars 
with similar quality spectra from the same program;
however, for the purpose of visual comparison to the Solar atlas, 
I chose to focus on those most similar to the Sun (i.e., $\teff < 6000$ K).}

I downloaded from the ESO Science Archive\footnote{\url{http://archive.eso.org/wdb/wdb/adp/phase3_main/form}} 
the \citet{Pace2004} data (66.D-0457(A), PI Pasquini),
which were obtained with the UVES spectrograph at the VLT Kueyen telescope, 
and which \citeauthor{Pace2004} reduced with the UVES pipeline \citep{UVESpipe}. 
The resolution of the \caiihk\ spectra are $R \approx 60,000$. 
The reader may refer to \citet{Pace2004} for additional details.

I supplemented the \citet{Pace2004} \caiihk\ spectra 
with 91 additional UVES spectra encompassing \nad\ for 30 unique M67 members, 
which were observed in order to measure 
beryllium abundances needed to study mixing in the stellar interiors 
of main-sequence FG stars (PI Randich, 69.D-0454(A), 68.D-0491(A), 65.L-0427(A))
and red giants (PI Duncan, 70.D-0421(A)).
This sample includes the solar analog S969, $(B-V)_0 = 0.62$, 
which was observed at $S/N \approx 55$ on 16 April 2000 
as part of 65.L-0427(A).

I also use a spectrum of the M67 blue straggler known as 
Fagerholm 280 \citep[F280, Sanders 1434, S1434; 
\teff\ = 9000 K, \vsini\ = 150 \kms ;][]{Milone1991},
which was acquired with the Tillinghast Reflector Echelle Spectrograph 
\citep[TRES, 3860--9100~\AA ;][]{andytres}
at the Fred Lawrence Whipple Observatory on Mount Hopkins, AZ,  
and kindly provided by Dave Latham (Harvard--Smithsonian Center for Astrophysics). 
G\'{a}bor F\H{u}r\'{e}sz's Ph.D. dissertation 
(2008)\footnote{http://doktori.bibl.u-szeged.hu/1135/}
describes TRES, 
and the reduction procedure is 
discussed in Lars Buchhave's Ph.D. dissertation 
(2010).\footnote{Entitled, ``Detecting and Characterizing Transiting Extrasolar Planets'' 
and available as a PDF online at the Niels Bohr Institute.}  
\subsection{Finding the clouds with interstellar sodium} \label{s:nad}
\begin{figure}\begin{center}
\plotone{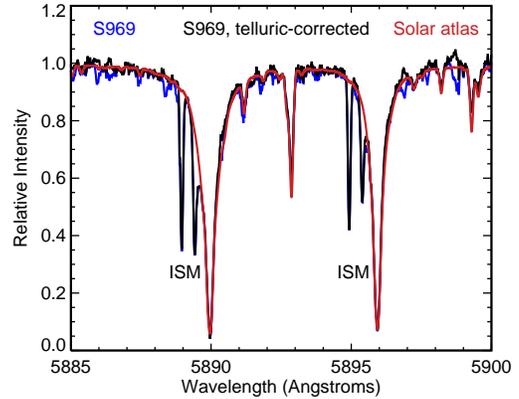}
	\caption{The Na D spectral region for M67 solar analog S969 
	(blue---observed, black---telluric absorption modeled and removed) 
	compared to the \citet{Wallace2011} solar atlas (red).
	The barycentric motion and radial velocity of S969 were subtracted to align it with the solar spectrum. 
    The solar spectrum matches the S969 spectrum quite well, 
    and reveals additional absorption lines that are formed in the interstellar medium. 
    \label{f:solar}}
\end{center}\end{figure}

At least three interstellar clouds occupy the foreground space 
between Earth and M67.
This is illustrated in Figure~\ref{f:solar}, which shows the 
\nad\ spectral region for the solar analog S969 with the Sun's spectrum overlaid. 
The solar atlas is a useful comparison because its stellar absorption features 
approximately match those seen in the S969 spectrum. 
This comparison highlights the interstellar lines and demonstrates the lack 
of prominent stellar features overlapping these ISM lines, 
aside from the wings of \nad , of course.
Note the lack of telluric sodium emission, 
which demonstrates that the sky subtraction procedure has 
effectively removed these strong lines.

\begin{figure}\begin{center}
\plotone{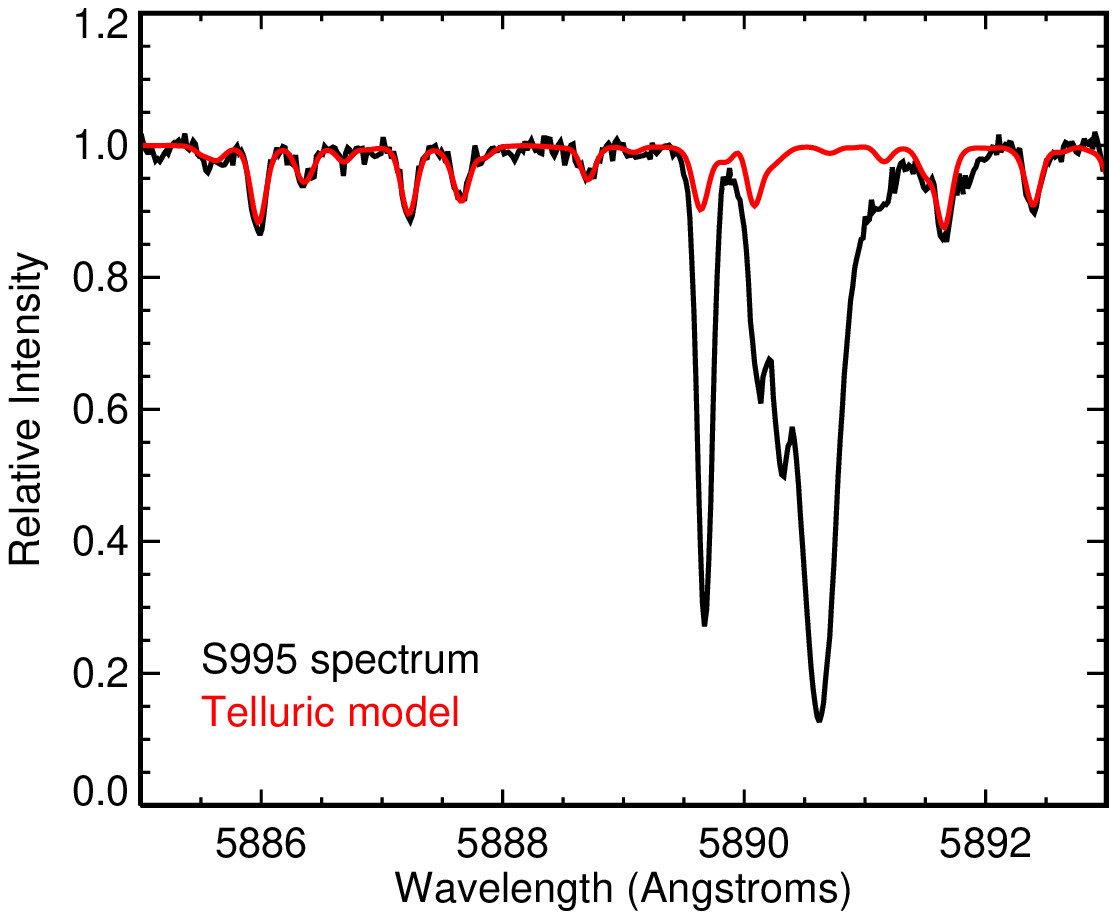}
\plotone{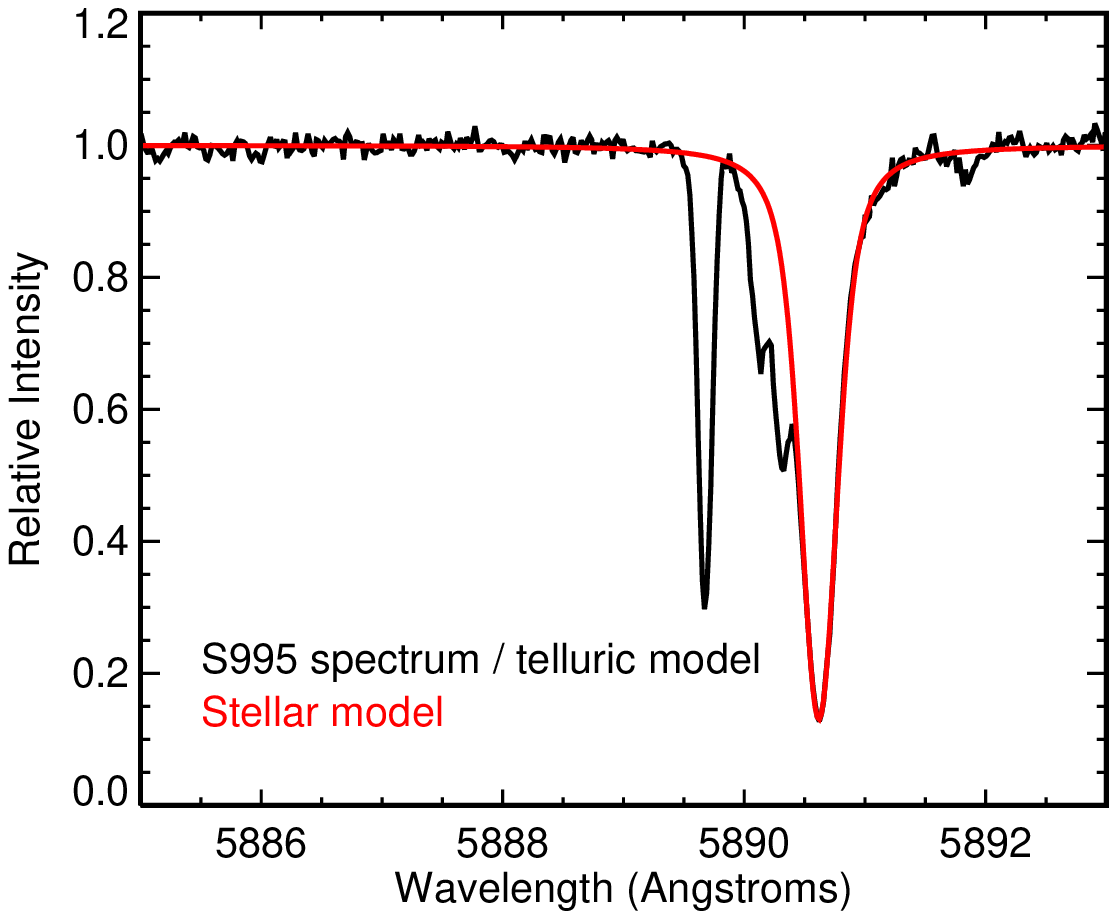}
\plotone{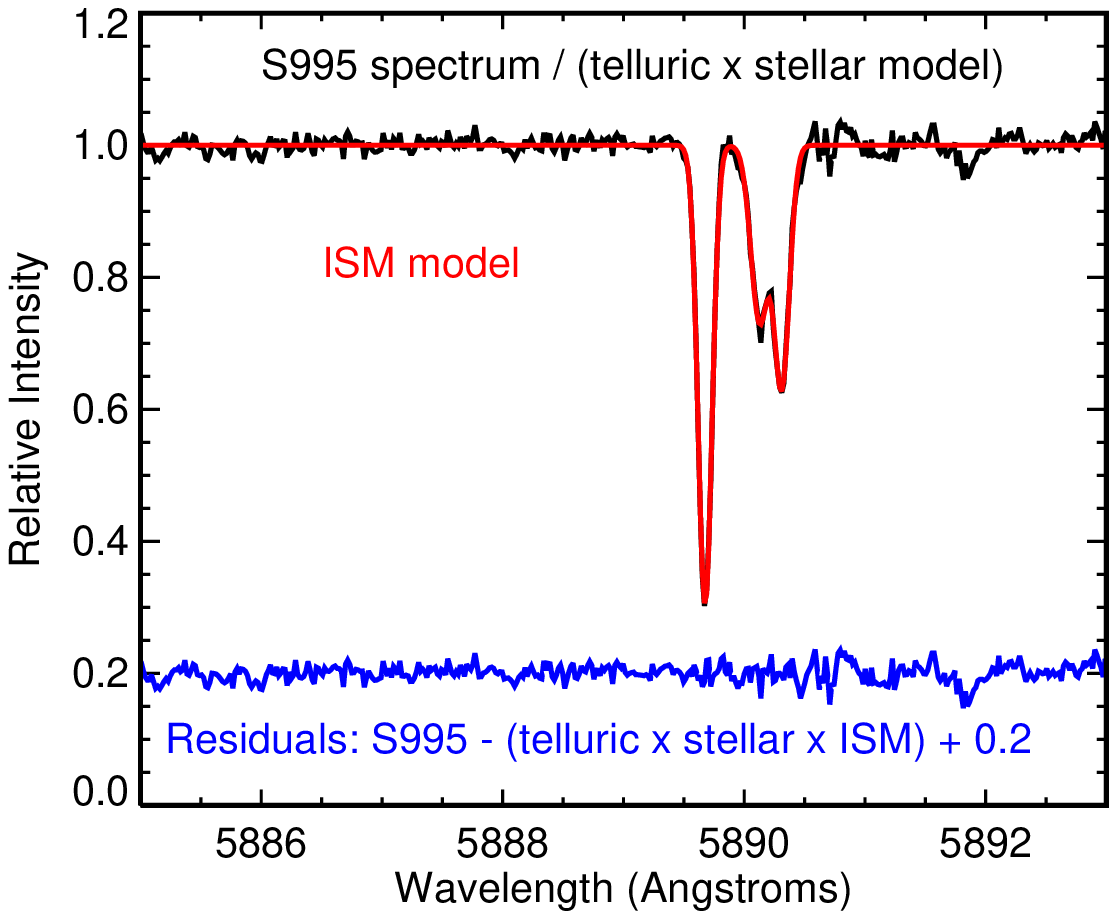}
	\caption{Illustration of the steps in the ISM-fitting procedure, 
	including telluric and stellar modeling.
	The Na D spectral region for S995, 
	a single member near the top of the main-sequence turnoff.
	The top panel overlays the \citet{Wallace2011} telluric spectrum in red, 
	which has been scaled by a constant multiplicative factor
    and smoothed with a Gaussian profile 
    to match the observed telluric line depths and profiles. 
	The middle panel shows the telluric-corrected spectrum, along with a 
	Voigt function that matches the stellar \nad 2 line profile. 
	The bottom panel demonstrates that the interstellar absorption can 
	be accurately described by a three-component Gaussian, shown in red, 
	which I derived by simultaneously fitting the telluric, stellar, 
	and interstellar spectral features, along with a linear normalization, 
	which has been applied to all panels. 
	The fitting residuals are shown in the bottom panel in blue, 
	shifted 0.2 upward.
	\label{f:NaDmethod}}
\end{center}\end{figure}

The \nad\ spectral region contains numerous telluric absorption lines
and proper care must be taken when modeling these features, 
along with the stellar spectra, in order to isolate the ISM lines. 
One can approach these issues in a stepwise manner; 
however, I found it is easier to simultaneously fit the telluric, 
stellar, and interstellar features, along with a linear normalization. 
Figure~\ref{f:NaDmethod} demonstrates the stepwise procedure 
in order to highlight the relative importance and structure of each component. 
The top panel plots the observed spectrum of S995 
(a single member near the top of the main-sequence turnoff), 
which has been normalized using the linear correction derived in the simultaneous fit; 
the red spectrum shows the \citet{Wallace2011} telluric spectrum, 
which has been scaled by a constant multiplicative factor
and smoothed with a Gaussian profile 
to match the observed telluric line depths and profiles. 
The middle panel plots the observed spectrum divided by the telluric model, 
along with a Voigt profile in red that fits the stellar \nad\ line. 
The bottom panel plots the observed spectrum divided by the telluric and stellar models; 
the interstellar clouds are accurately described by a 3-component Gaussian model shown in red, 
with the residuals shown in blue and shifted upward by 0.2 in relative intensity for 
ease of visualization. 
This spectral region can therefore be described by 17 parameters: 
the linear normalization (2: slope and intercept), 
a telluric model (2: broadening and line depth scaling applied to the 
\citeauthor{Wallace2011}~\citeyear{Wallace2011} telluric spectrum), 
a stellar \nad\ profile (four-parameter Voigt function, normalization to unity assumed), 
and a three-component Gaussian for the ISM (nine parameters for the centroid, width, depth for each, 
where normalization to unity is again assumed). 
Fitting these 17 parameters simultaneously is straightforward because they are generally independent, 
where changing one parameter has little effect on other parameters (e.g., the telluric parameters are 
determined from spectral regions distinct from the stellar and ISM lines; the stellar and interstellar 
velocities are distinct).

\begin{figure}\begin{center}
\plotone{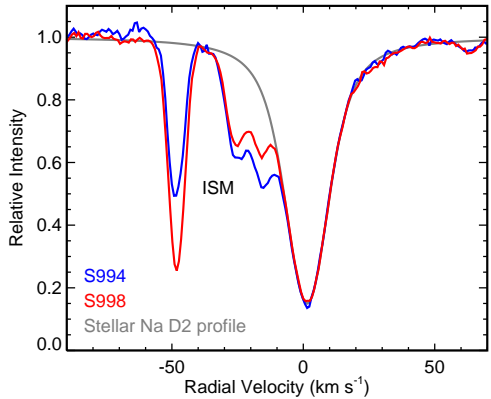}
	\caption{ Na D spectra for analogous members of M67 show differential ISM line strengths.
	Matching analogs like this 
	is useful for isolating the ISM lines, and I find that 
	the strengths of the three ISM lines vary depending on the 
	particular sight line, suggestive of differential extinction, 
	and necessitating star-by-star ISM corrections for the activity 
	measurements described in later sections. 
	The barycentric motions and cluster RVs have been subtracted from these spectra. 
	\label{f:NaDiff}}
\end{center}\end{figure}

I applied the simultaneous fitting procedure to all 91 UVES \nad\ spectra of the 30 unique stars 
across the cluster, and I see clear evidence for differential ISM absorption. 
This is readily demonstrated by plotting the spectra for pairs of analogous members 
(i.e., similar spectral types in different parts of the cluster that were selected 
according to similar optical colors and magnitudes), 
for which I provide an example of in Figure~\ref{f:NaDiff}.
Three features are clearly seen toward every star I tested, 
each of which might be comprised of a single cloud or multiple clouds unresolved at this spectral resolution. 
The interstellar absorption lines appear to vary in strength 
from star to star. 

\subsection{Modeling the clouds with blue stragglers}\label{s:model}

Although the interstellar \nad\ lines can be modeled with the straightforward 
procedure outlined above, the \caiihk\ lines are more difficult for most stars. 
First, it is challenging to obtain high-resolution spectroscopy 
at sufficiently high $S/N$ for the main-sequence GK dwarfs in 
M67 or the \textit{Kepler} clusters (i.e., 1 Gyr NGC 6811 and 2.5 Gyr NGC 6819) 
due to their large distance moduli. 
Again, this difficulty is compounded when observing \caiihk , 
because the chromospheric emission and the contaminating ISM lines are found 
in the deep line cores. 

It can be additionally difficult to model the ISM lines toward GK dwarfs when 
the relative velocities of the clouds and stellar line core emission are 
coincident because we cannot know the stellar emission profile a priori 
in order to model the ISM line. This is why we care to observe these stars, after all!
Cluster blue stragglers are useful alternatives because they tend to be hotter and 
rotate more rapidly than the old, cool GK dwarfs of interest.
Rotational broadening in particular can dramatically simplify the \caiihk\ region, 
leaving interstellar lines standing in sharp contrast to the few remaining smooth and broad stellar lines.

\begin{figure*}\begin{center}
\plottwo{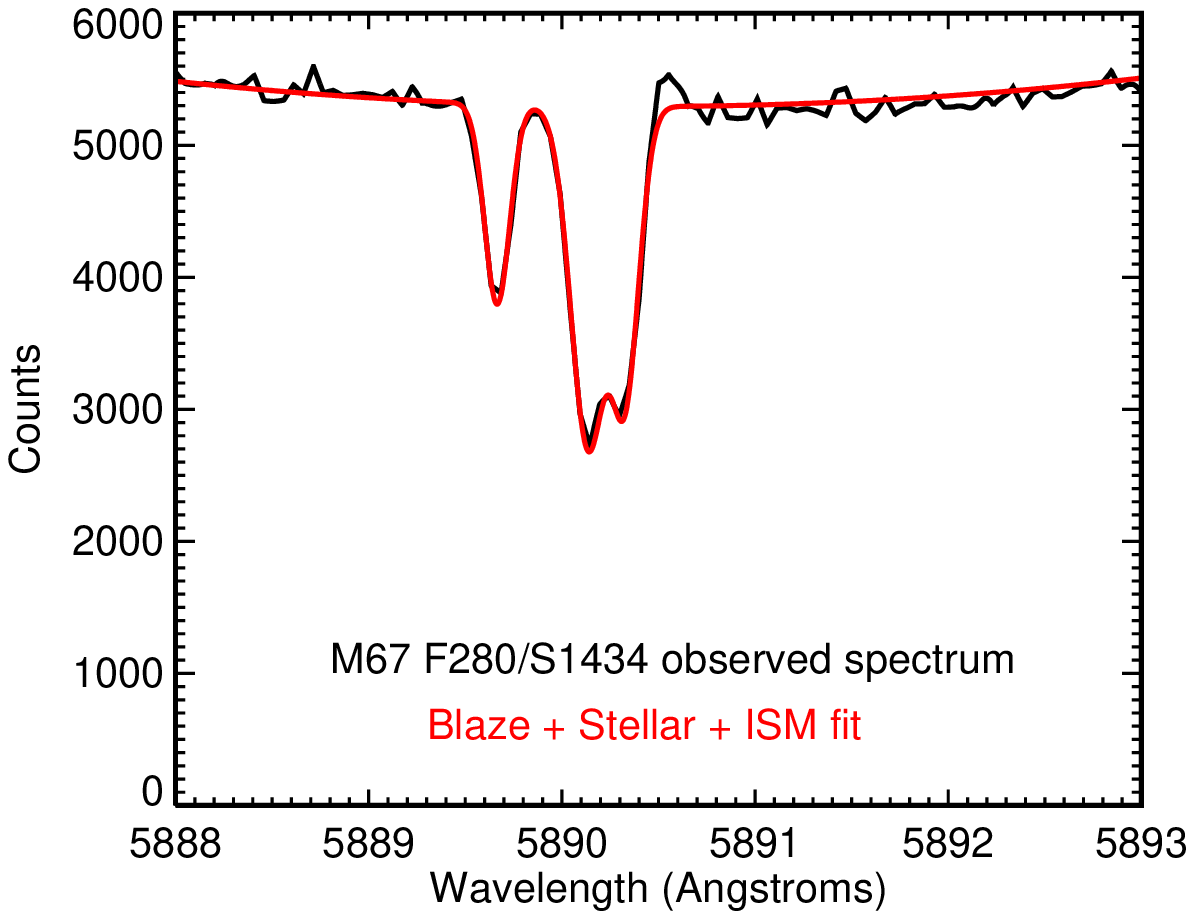}{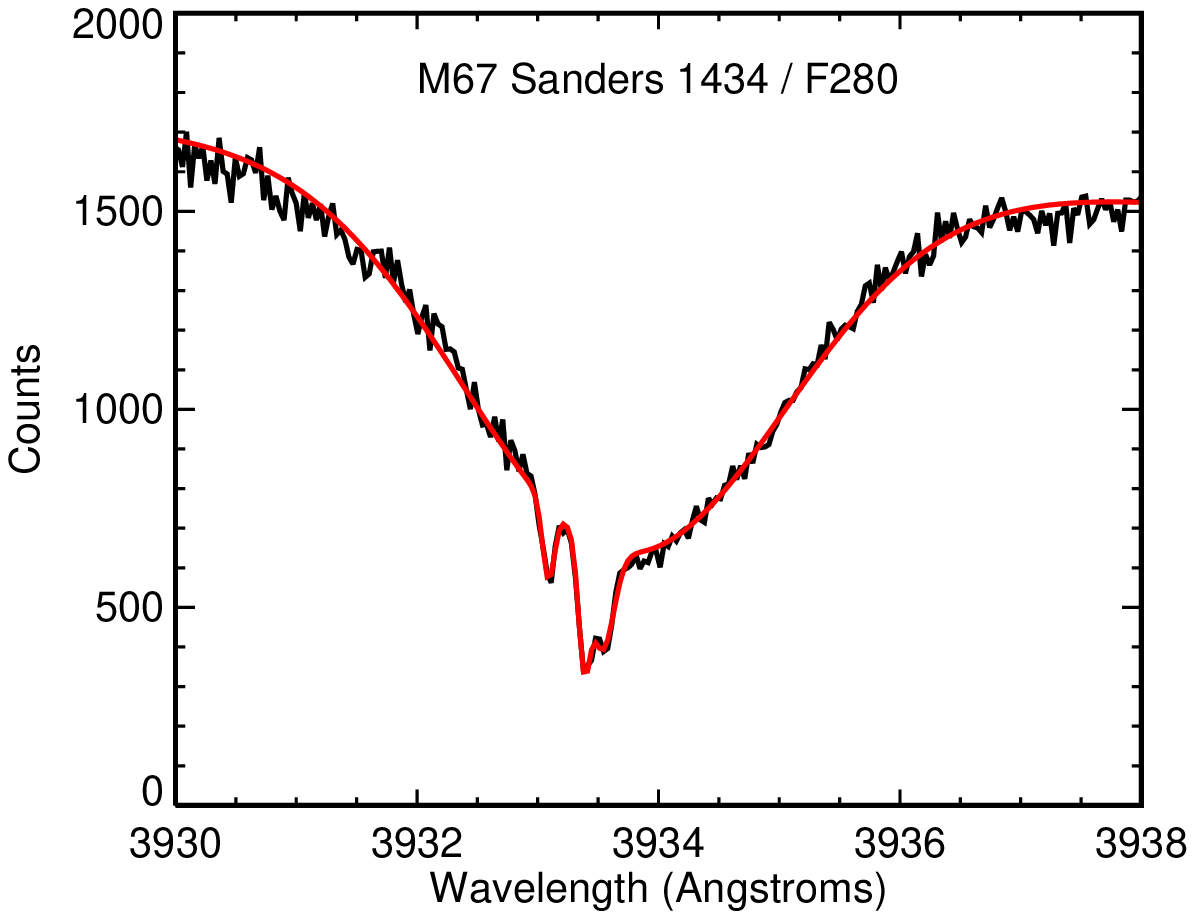}
\plottwo{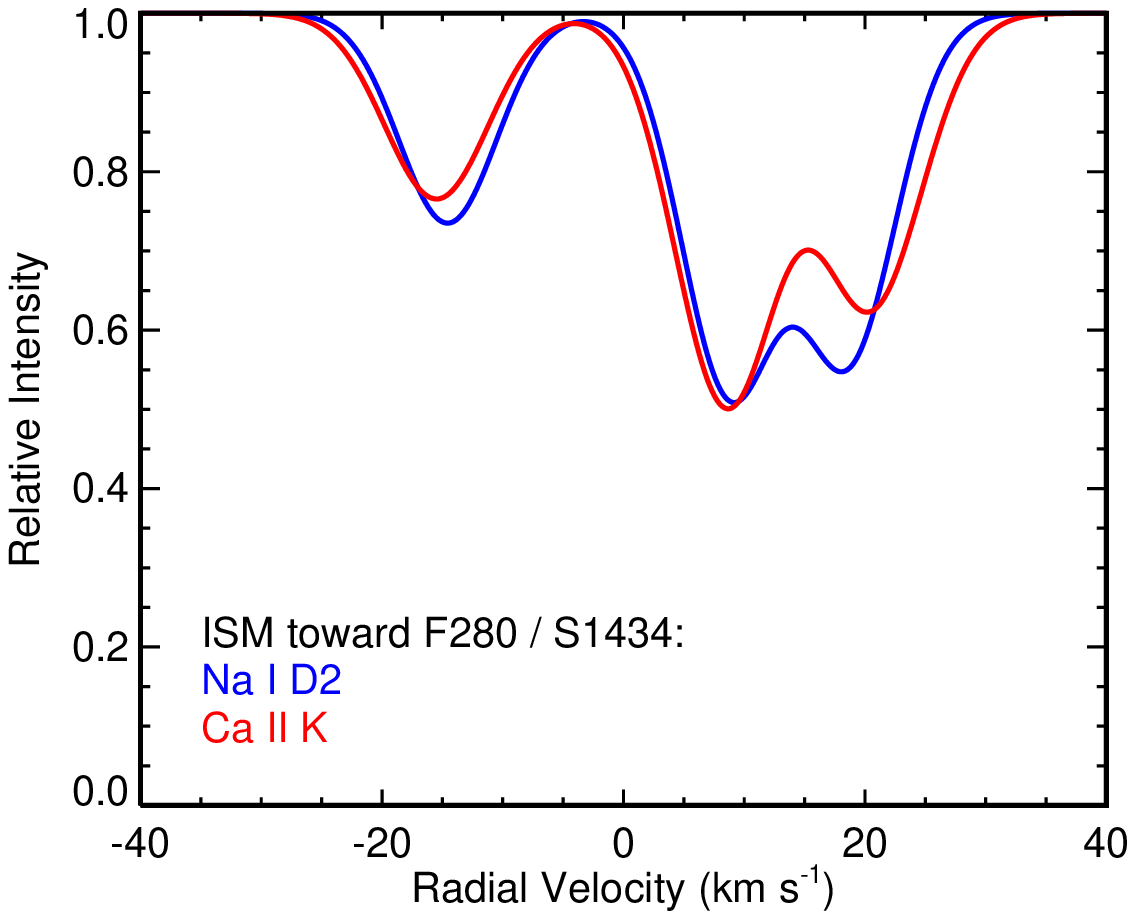}{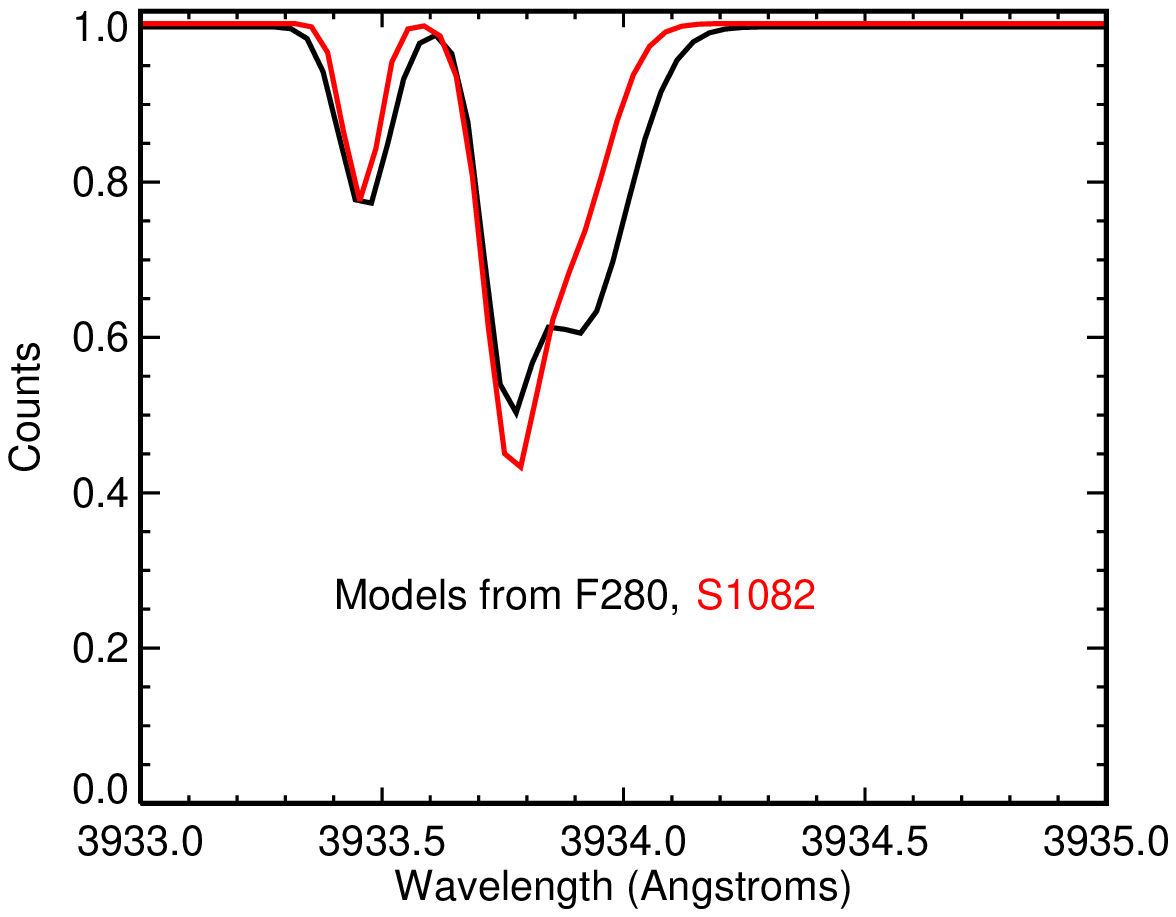}
	\caption{Deriving ISM models for Na D and \caii\ lines from blue straggler spectra:
	The top-left and top-right panels show 
	\nad 2 and  \caii\ K TRES spectra for the blue straggler F280 (S1434), 
	along with my best-fitting models. 
	The bottom-left panel overlays the resulting three-component Gaussian model 
	of the ISM fit from the \nad 2 (blue) and \caii\ K (red) lines.
	The bottom-right panel compares the \caii\ K ISM models derived 
	from S1434 (black) and S1082 (red), which are separated by 15 arcminutes
	(projected separation of 4 pc at a distance of 880 pc).
	The difference appears to support the presence of differential column depth. 
	\label{f:bss}}
\end{center}\end{figure*}

The top panels of Figure~\ref{f:bss} show \nad 2 and \caii\ K regions for the 
M67 blue straggler F280/S1434. 
For the blue stragglers F280/S1434 and S1082, 
I fit the \caiihk\ regions simultaneously with a three-component Gaussian to account for the ISM, 
with RVs and line depths for each component modeled by the same parameter between H \& K; 
the line depths for \caii\ K are multiplied by two relative to the H lines to account 
for the difference in strengths between the two lines.\footnote{This is due to 
the degeneracy of the \caii\ K line transition, which 
originates from the $J = 3/2$ state, where $g = 4$, while 
the \caii\ H line is from the $J = 1/2$ state with $g = 2$ \citep[page 93 of][]{Tennyson2005}.}
The broad stellar H \& K lines and blaze functions were fit independently 
with Gaussians to approximately match the stellar profiles and 
quadratic polynomials to match the local blaze functions. 

The bottom-left panel overlays the ISM models derived from the Na D and \caiihk\ 
spectra. 
The ISM features at approximately $-49$ and $-25$ \kms\ appear strikingly similar, 
while the velocities for the $-13$ \kms\ cloud are discrepant by $\sim$3 \kms . 
In the future, I will incorporate multiple observations of this star to increase 
the $S/N$ and improve the quality of the ISM model, 
hopefully bringing the \caiihk\ and \nad\ ISM models into better agreement.

The bottom-right panel overlays the ISM models for F280/S1434 and S1082. 
Again, we see non-negligible differences in interstellar absorption between cluster members.
Table \ref{t:ism} provides the results of the Gaussian fits 
for the interstellar features for the blue straggler F280/S1434. 

\begin{deluxetable}{cccc}
\tablecaption{\caii\ K ISM absorption model for M67.  \label{t:ism}}
\tablewidth{0pt}
\tablehead{
\colhead{Sanders ID} & \colhead{Component} & 
\colhead{RV} & \colhead{Depth} \\
\colhead{} & \colhead{} & 
\colhead{(\kms )} & \colhead{}
}
\startdata
1434 & 1 &  $-12.67$ & 0.397 \\
...  & 2 &  $-24.91$ & 0.508 \\       
...  & 3 &  $-48.77$ & 0.232 \\
\enddata
\tablecomments{(1) Sanders ID (star resolvable with SIMBAD as ``Cl* NGC 2682 SAND 1434''),
(2) component number, starting with the nearest to \caii~K at 
3933.66~\AA , 
(3) radial velocity of ISM component. 
To apply the model to \caii~H, apply these RVs to 
its rest wavelength at 3968.47~\AA\ 
(4) line depths for Gaussian models normalized to unity. 
To compute equivalent widths or generate models, 
the standard deviation of the Gaussians should be 0.06~\AA . 
The Gaussian widths should be adjusted for different spectral 
resolutions, and the line depth adjusted accordingly 
to preserve equivalent width. 
For \caii~H, divide the line depths by two. 
} 
\end{deluxetable}

\subsection{Validating the model with M67's solar analogs}\label{s:validation}
\begin{figure}\begin{center}
\plotone{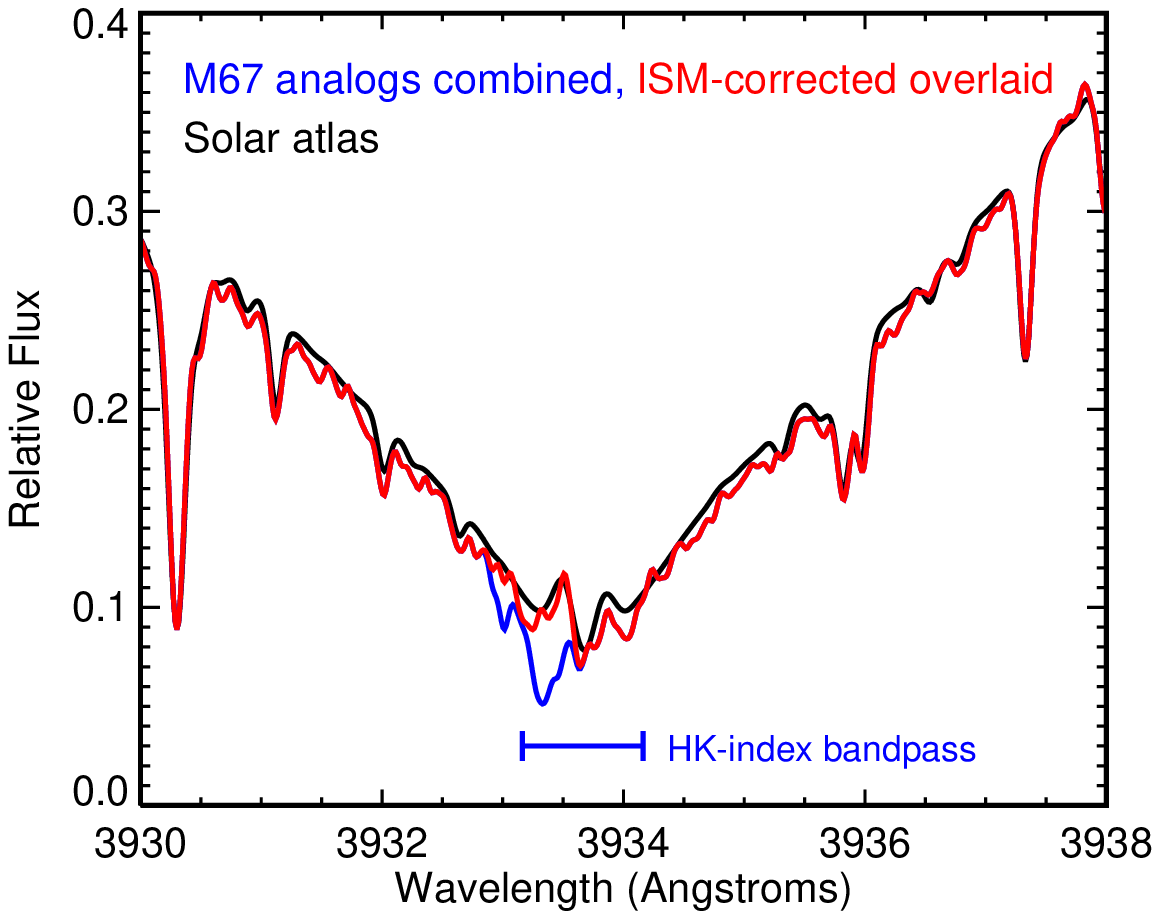}
\plotone{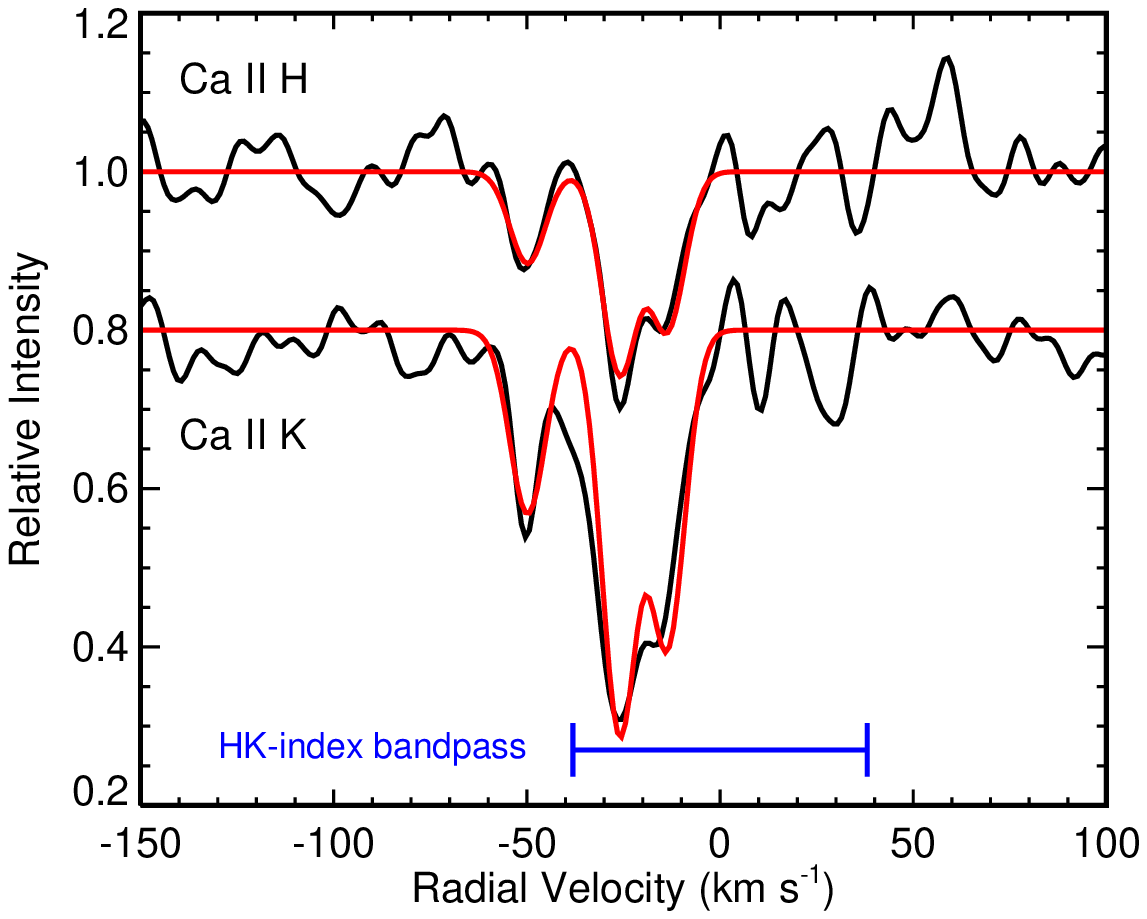}
	\caption{Comparison of VLT UVES spectra of M67 solar analogs and the Sun 
	validates the blue straggler ISM model:
	Top---A \caii\ K spectrum of a typical Solar analog in M67. 
	        Six VLT UVES spectra for three individual stars were co-added to produce 
	        a cycle-averaged spectrum with sufficient $S/N$ to validate the 
	        ISM model. The red plot shows the co-added spectrum divided by the 
	        ISM model derived from the blue straggler F280;
	        The uncorrected spectrum is underlaid in blue. 
	        The \citet{Wallace2011} solar atlas is underlaid in black, 
	        and closely matches the stellar lines and the chromospheric emission profile.
	        As you can see, the ISM model restores symmetry to the \caii\ K line core profile. 
	 Bottom---The co-added M67 spectrum divided by the Solar atlas. 
	        Given the quality of the co-added spectrum and the similarity in spectral type with the Sun, 
	        this ratio yields an additional empirical model for the ISM features. 
	        The blue straggler model is overlaid in red and produces an accurate match 
	        despite the apparent ISM-variability noted earlier.
	\label{f:M67sun}}
\end{center}\end{figure}

While the $S/N$ of the VLT/UVES \caiihk\ spectra for each individual solar analog 
is insufficient for modeling the ISM line ($S/N \approx 10 - 20$), 
co-adding six spectra for three single solar analog members 
(Sanders 746, 1048, 1255) 
with barycentric motions removed
yielded a spectrum of decent quality that
(1) revealed the ISM features and 
(2) averaged the \caiihk\ line core emissions, effectively averaging over cycle and short-term
variability, to produce a \caiihk\ spectrum of a typical M67 solar analog. 
It was necessary to select single (i.e., non-binary) stars to preserve the relative 
RVs of the ISM clouds and stars; this also rejected the overactive binary S1012. 
The top panel of Figure~\ref{f:M67sun} shows the co-added M67 solar analog spectrum in blue, 
with the same spectrum divided by the ISM model overlaid in red---the 
only portion of the original spectrum still remaining is from the ISM, 
which was apparently well-modeled from the BSS.
I also overlaid the solar atlas, which nicely traces each stellar line, 
and happens to closely match the \caiihk\ core emission as well. 
Introducing the ISM correction restores 
symmetry to the \caii\ K line core profile. 
Given the close match between the Solar atlas and the M67 co-added spectrum, 
dividing the two inverts the problem and yields an additional empirical model 
for the ISM features, which is shown in the bottom panel of Figure~\ref{f:M67sun}.
The ISM model derived from the blue straggler 
accurately matches the averaged ISM toward these three solar analogs.

\section{Mitigating ISM-contamination for HK} \label{s:unbias}
With an accurate ISM model in-hand, I will now calculate the magnitude of 
the negative bias. 
When applying this model to individual stars, it is important to account for
the barycentric motion for the particular observation, 
and crucial to adjust the ISM RVs for
any additional relative motion between the star and the
cluster's bulk motion; for example, if the system is a
spectroscopic binary.

\citet{Giampapa2006} measured chromospheric activity with the HK index, 
which integrates the emission in 1~\AA\ windows centered on the \caiihk\ line cores.
I refer the reader to that work and references therein for a discussion of 
spectrum normalization and index calibration \citep[particularly][]{HallLockwood1995}. 

\begin{figure}\begin{center}
\plotone{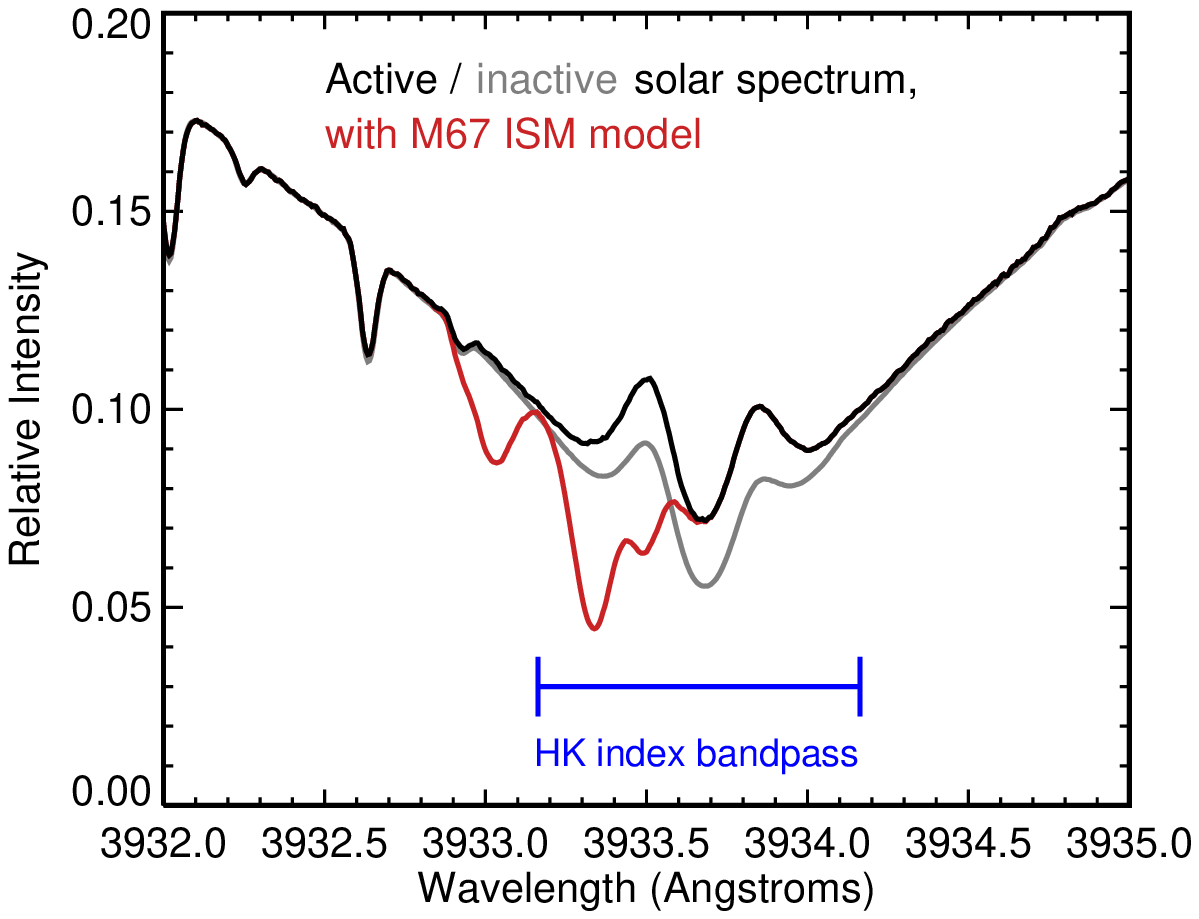}
\plotone{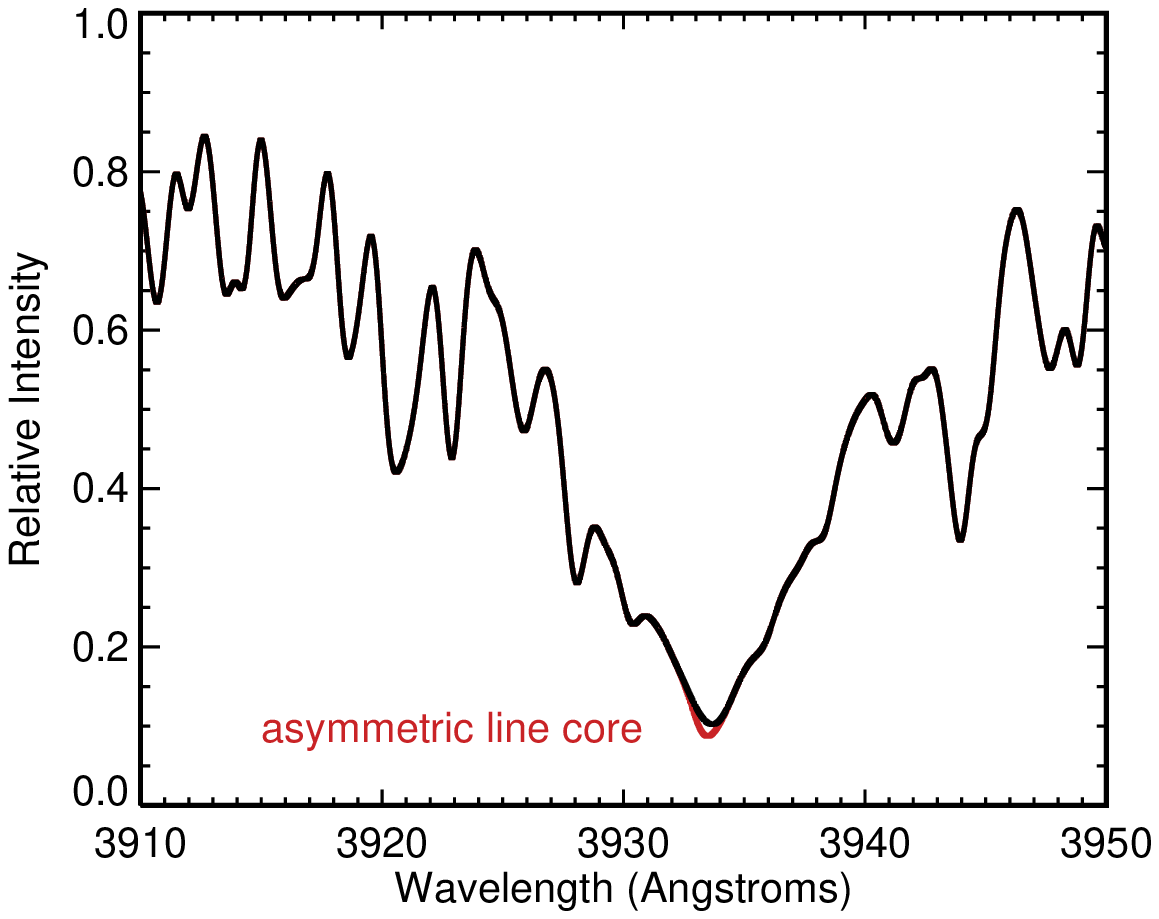}
	\caption{M67 ISM models are applied to solar spectra to 
	derive HK-index corrections.
	Top---A high-resolution solar spectrum from SOLIS/ISS
	(black---high activity, gray---low activity)
	has been modified according to the three-cloud M67 ISM model (red), and  
	the 1~\AA\  K index bandpass is marked for reference.
	The median ISM bias for the integrated H \& K index is 12.3 and 6.7~\ma , 
	for a total of 19.1~\ma\ at high spectral resolution.
	Bottom---The \citet{Wallace2011} solar atlas has been ``degraded'' to a spectral resolution 
    of $R = 5000$, similar to the WIYN/Hydra instrumental setup used by \citet{Giampapa2006}.
    This panel is shown zoomed out relative to the left panel to facilitate visual comparison to Figure~2 
    from \citet{Giampapa2006}:  we note an asymmetric line core deficit for the M67 ISM model 
    spectrum here, which appears similar to the plot of the Sun and M67 S1246 
    \caiihk\ spectra in that figure. 
    At this lower resolution, the ISM negatively biases the H \& K indices downward by 5.3 and 9.7 \ma , 
    for a total of 14.9~\ma , 
    which is approximately equal to the maximum activity deficit found 
    by \citet{Giampapa2006} for 15\% of the Sun-like stars.
	\label{f:sun}}
\end{center}\end{figure}

I used the SOLIS/ISS \caiihk\ spectral library and applied the ISM model derived 
from the blue straggler F280/S1434 to simulate M67 solar analogs with activity levels 
spanning the contemporary Sun's as captured by SOLIS between 2006 December and present. 
This is illustrated in the top panel of Figure~\ref{f:sun}. 
I computed the HK index for both the 
ISM-contaminated and ISM-free spectra at high resolution. 
Next, I broadened these spectra with a Gaussian kernel of FWHM = 0.8~\AA\
to approximately match the spectral resolution of the Hydra observations and 
re-calculated the indices.
In units of \ma\ and quoting the low-resolution results in parentheses, 
I find that the ISM model negatively biases the K index by 12.3 (9.7) and 
the H index by 6.7 (5.3), 
giving a total correction of 19.1 (14.9). 

The bottom panel of Figure~\ref{f:sun} shows the \citet{Wallace2011} atlas  
broadened with the same Gaussian kernel to mimic the Hydra observations.
Note the asymmetric line core of the simulated M67 spectrum. 
This shape is similar, and this figure is analogous, to Figure~2 from \citet{Giampapa2006}, 
which compared the solar spectrum with Sanders 1246, a $(B-V)_0 = 0.64$ solar analog.
There, it simply appeared that S1246 was less active than the solar epoch 
(HK = 187 m\AA , compared to the Sun's 68\% range of 188 to 198~m\AA ); 
however, that plot was actually showing an asymmetric line core profile 
caused by the unresolved presence of interstellar \caii . 

To derive accurate and precise indices, the ISM model should be applied directly to the Hydra 
spectra because the correction value is somewhat dependent on spectral type 
(the \citeauthor{Giampapa2006}~\citeyear{Giampapa2006} 
sample includes stars with $(B-V) = 0.51-0.79)$, 
activity level, and of course, stellar radial velocity. 

\section{No Maunder minimum candidates in M67}
Early efforts to identify MM candidates in the solar neighborhood 
were informed by the observation of very inactive stars 
with flat records \citep[e.g., $\lrphk < -5.1$~dex;][]{Baliunas1990, Henry1996}.\footnote{Converting
$\lrphk = -5.1$~dex to HK is $\approx 160$~\ma\ on the \citet{Giampapa2006} scale.}
Since those stars turned out to be subgiants \citep{wright_maunder}, 
it is important to return to the defining characteristics of Maunder-like grand minima. 
During the Maunder Minimum, the Sun must certainly have exhibited reduced chromospheric heating and emission  relative to the contemporary Sun 
due to the direct correlation between sunspot number and \caiihk\ emission 
observed in the modern era \citep{White1981}.
Whether this reduced activity level was on par with the modern solar minimum or 
if it was in deficit to the minimum is not known \citep[e.g., approaching or equal to the basal photospheric emission 
that has been measured from solar center-disk observations, 
described as representing ``immaculate photospheres'' by][]{Livingston2007}.
The $^{10}$Be radionuclide record modulated at the $\sim$11~year solar cycle period 
throughout the Maunder Minimum, so apparently the solar magnetic dynamo had not shut down completely \citep{Beer1998}.
As \citet{Beer1998} explained, 
there must have been coronal heating in order to drive the wind that carried the solar fields that 
would have deflected the Galactic cosmic rays that produced $^{10}$Be in Earth's atmosphere 
to varying degrees throughout the weakened solar cycle.
Whatever magnetic activity persisted probably produced weak HK emission in excess to the center-disk 
level. \citet{Livingston2007} quote a center-disk HK index of $150 \pm 7$~\ma , 
or 157~\ma\ on the \citet{Giampapa2006} scale. 

Identifying MM candidates solely according to their \caii\ emission is challenging because, 
in addition to not knowing how active the Sun really was during the Maunder Minimum, 
our activity indices (e.g., \rphk ) are not calibrated to account for 
metallicity (e.g., line blanketing in the pseudo-continuum) and surface gravity. 
As \citet{Hall2004} and \citet{Saar2006} state, 
we cannot simply adopt a threshold activity for MM candidates in the field 
\citep[see also][]{Hall2009}.
Instead, \citet{Saar2012} fitted the lower envelope of activity in the joint 
\citet{wright2004}--\citet{valenti2005} dwarf sample as a function of metallicity 
and identified candidates that fell below this line 
and exhibited small dispersions (i.e., approximately flat activity).

\citet{Judge2007} and \citet{Saar2012} examined the X-ray luminosities and UV emission line fluxes 
of MM candidates (e.g., $\tau$~Ceti, also known as HD~10700) 
and found levels similar to the cycle-minimum Sun. 
If those stars are truly analogous to the Sun during the Maunder Minimum, 
then this would suggest that the solar chromosphere, transition region, and corona 
were still magnetically heated even while sunspot production in the photosphere was significantly reduced. 

Surveying M67 (and solar twins in the field) addresses the metallicity/HK-calibration issue 
by focusing on stars with bulk properties (e.g., mass, \teff , \logg , \feh ) 
very similar to the Sun, 
since differentially comparing activity indices between the Sun and solar twins will be unaffected 
by differences in surface gravity and metallicity.\footnote{I believe this strategy 
of focusing on solar twins in the 
field and nearby clusters should be pursued in parallel to, not in lieu of, observations 
of bright, non-solar metallicity FGK dwarfs that were targeted by the Mount Wilson survey or 
are being monitored with the Solar--Stellar Spectrograph at Lowell Observatory in Flagstaff, Arizona
\citep{HallLockwood1995, Hall2004}.}
It is possible that the difference in HK emission during regular cycle minima and 
the Maunder Minimum might be so subtle as to be indistinguishable, 
which would make HK a poor or ineffective indicator of ``Maunder-hood.'' 
Furthermore, given that the Sun apparently maintained a reduced dynamo cycle throughout the Maunder Minimum 
\citep{Beer1998}, this places even greater importance on checking the HK histories, albeit limited 
to a handful of epochs over a period of six years, for evidence of variability typical of a normal 
solar cycle to rule out ``Maunder-hood'' for stars with average HK levels near or below solar minimum. 
Therefore, the two key criteria I adopt for evaluating a star's ``Maunder-hood'' in this study
are a relatively flat activity record lasting longer than a typical cycle minimum 
(or perhaps one that transitions into or out of an extended flat state)
and chromospheric/coronal emission near or perhaps below the modern solar minimum level. 

Again, I will classify M67 members as overactive or underactive as those departing from 
the solar range of 179 to 226~\ma . 
In light of \citet{Judge2007}, I will also consider stars near solar minimum.
According to the monthly averaged series, the Sun spent 95\% of the past 110 years between 183 and 209~\ma . Therefore, I will also review the activity time series plots \citep[Figure 9 by][]{Giampapa2006} 
for stars within 5~\ma\ of solar minimum (i.e., $<185$~\ma ) to check for variability or a flat record.

\subsection{The M67 HK distribution}
\begin{figure*}\begin{center}
\plottwo{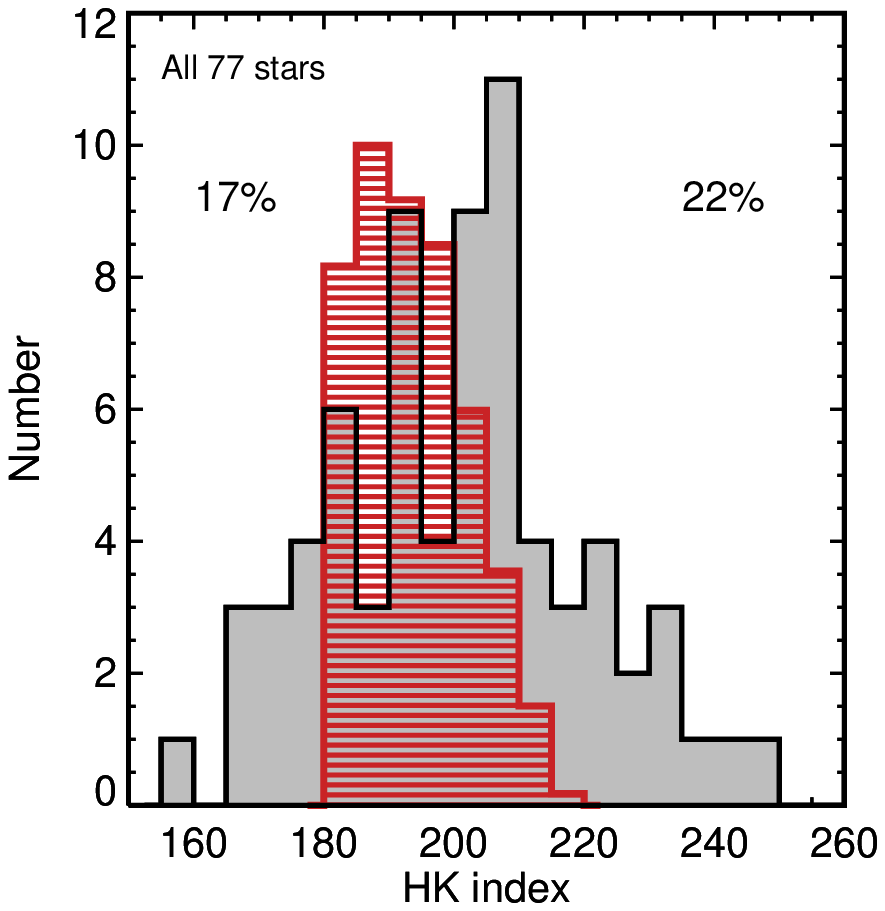}{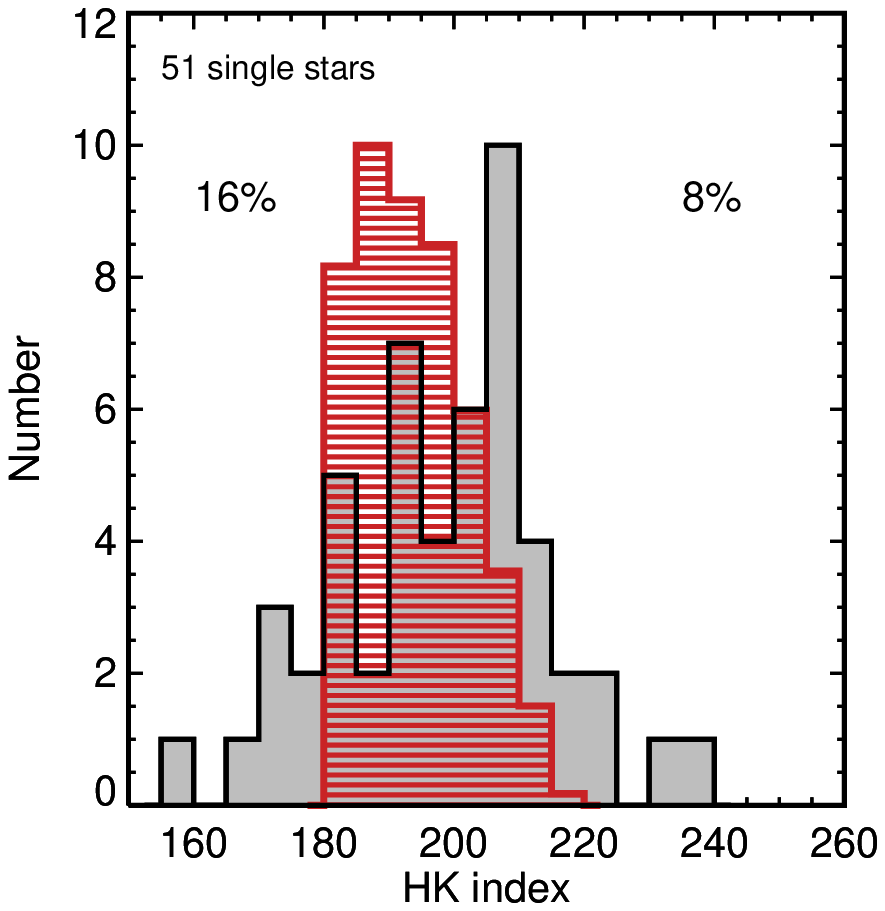}
\plottwo{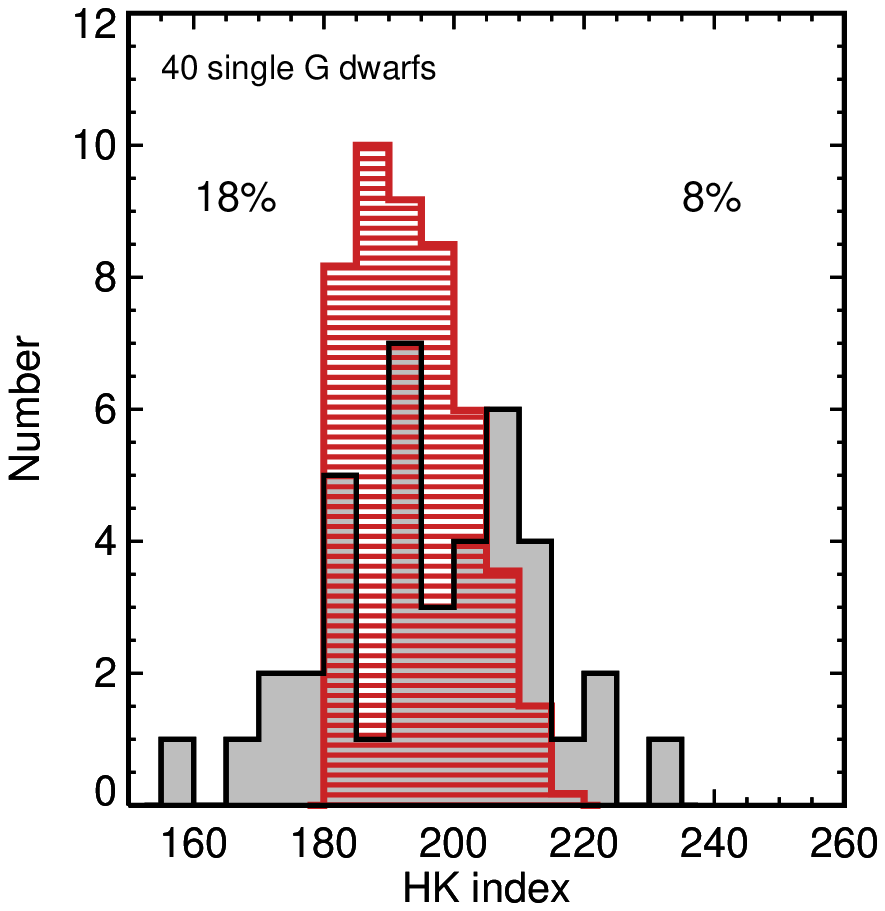}{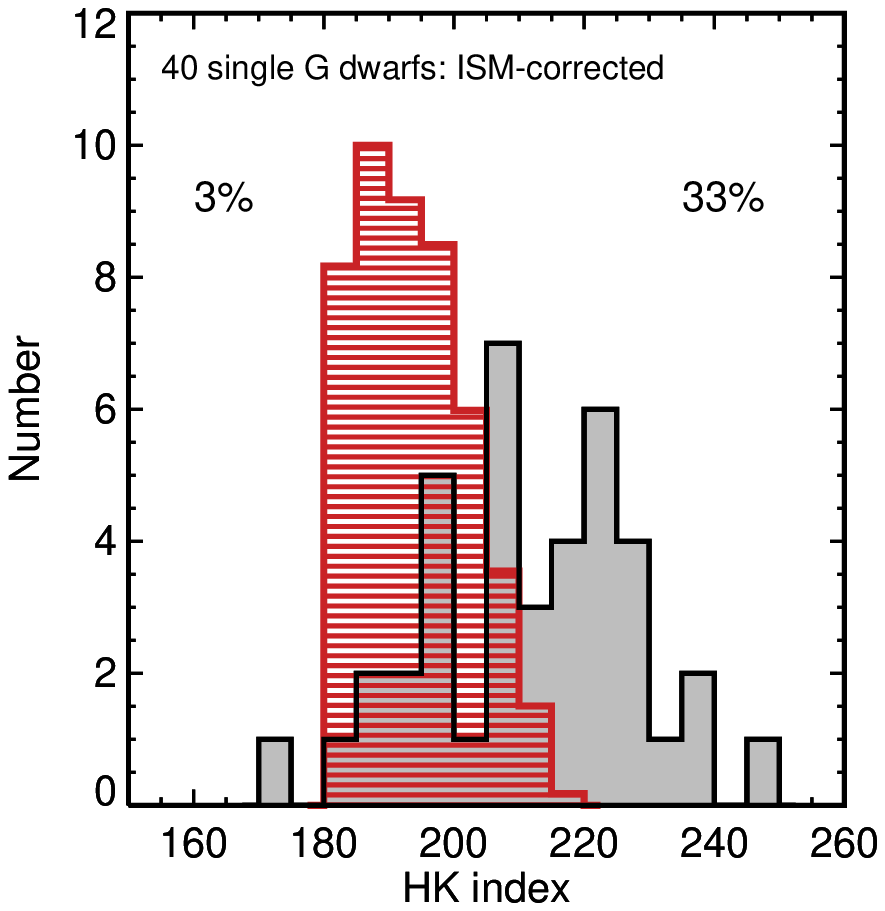}
    \caption{
    M67 HK-index distribution (gray) for 
    the full sample (top left), 
    the single stars (top right), 
    the single G dwarfs (bottom left), 
    and the same single G dwarfs with HK corrected for the ISM bias (bottom right), 
    compared to the 100 year solar history (red, normalized to $N=10$).
    Top left---The distribution of uncorrected HK from \citet{Giampapa2006} for 
    all stars that are classified as members by \citet{Geller2015}. 
    The typical solar range is marked by red dashed lines. 
    Of the 77 stars with HK indices, 73 are classified as members 
    (50 single and 19 binary members, and 4 binary likely members);
    14\% appear less active than solar minimum and 16\% are in excess of solar maximum. 
    Top right---This panel shows the HK distribution for the 50 single members, 
    14\% of which are below Solar minimum; 
    however, only 4\% are in excess of solar maximum, 
    which suggests that the binaries might have either 
    experienced tidal interaction, 
    or the relative velocity of the primary might have moved 
    the ISM lines outside of the HK-bandpass, 
    which would raise the apparent activity level relative to the 
    single-star sample. 
    Bottom---This panel introduces the 22~\ma\ ISM correction 
    to the HK-indices for single members.
    Only one star is now found below solar minimum: S1473 
    was observed with HK = 154~\ma\ in 1997 and 188~\ma\ in 2002, 
    which suggests that it is not in a Maunder minimum state. 
    However, it is important to explore why S1473 reached 
    this low-activity level at another time. 
    Finally, the activity levels of 24\% of the single stars 
    now appear to be in excess of solar maximum, 
    and the mean level for the entire sample is higher 
    than the modern solar mean. 
    This is consistent with the model for continuous, 
    monotonic decay of chromospheric activity over time, 
    given that M67 is 500 Myr younger than the Sun, 
    according to best estimates. 
    However, the accuracy and precision of this age estimate 
    are not sufficient to tell if M67 is, in fact, 
    younger than the Sun. 
    Important results from \textit{K2} (e.g., red giant asteroseismology for improved age precision, 
    rotation period measurements for G dwarfs), along with an improved 
    distance modulus from high precision Gaia parallaxes, 
    will hopefully help clarify this issue. 
    }   
    \label{f:hk}
\end{center}
\end{figure*}

The distribution of HK activity measurements for 77 Sun-like stars from 
\citet{Giampapa2006}, which are plotted in the top-left panel of Figure~\ref{f:hk}, 
spans a greater range than experienced by the modern solar cycle.
Approximately 17\% of the stars showed mean activity levels 
below solar minimum, 
and 22\% of the stars were in excess of solar maximum. 
While \citet{Giampapa2006} did investigate the role of binarity
using R. D. Mathieu's membership catalog that was provided via private communication, 
additional binaries were identified in the intervening 10 years. 

Of the 77 stars in the \citet{Giampapa2006} sample, 
\citet{Geller2015} classifies 
50 as single members, 
 3 as a single non-members,
 1 as a binary non-member,
19 as binary members, and
 4 as binary likely members.
I argue that the four non-members are likely members and will treat them as such for 
this investigation 
(here I use RV and binary orbital solution information provided to me 
by D.~Latham):

\textit{S753} was classified as a single non-member despite its RV being constant and consistent 
with membership over 26 years due to a lack of proper motion information. 
\citet{Geller2015} used proper motions from four sources and required that 
the majority indicated membership (i.e., $P > 50\%$). 
In this case, three sources provided values---one gave 43\%, another 93\%, and a final 0\%.  
\citet{Geller2015} happened to place the greatest weight on the source 
that quoted the 93\% probability in cases of a tie; however, there was no tie here. 
Still, the first source did quote a non-zero probability 
and 
the other two sources suggest either definitely or definitely not a member. 
Perhaps in this case, proper motion should not be used to settle membership.
Given the long history of RV consistency, I conclude that S753 is a single member.

\textit{S1107} was also classified as a single non-member. 
D.~Latham observed this star 20 times over 6.19 years and measures a mean RV equal to 
the cluster velocity with rms of 2.2~\kms . 
The RVs appear to trace a sinusoid, suggesting that this is an SB1 with an approximately 6~year period. 
I re-classify it as a binary member.

\textit{S1203} was classified as a binary non-member. 
Latham finds that it is an SB1 with a 22.4~day period and a semi-amplitude of 13.9~\kms . 
Although it satisfies the proper motion criteria (probabilities from three sources are 92\%, 97\%, and 97\%), 
the center-of-mass RV is 4.5~\kms\ discrepant, 
whereas the cluster velocity dispersion is only 0.6~\kms . 
This could be simply explained by the presence of a long period tertiary. 
In fact, the RVs only span 3.3~years, which is likely not long enough to 
detect the acceleration from a third star. 
I re-classify this as a binary likely member under the assumption that there is (at least) 
a third star that is responsible for the modest RV offset.

\textit{S1246} was classified as a single non-member. 
Latham's RVs show a 5~\kms\ ``trend'' over 23 years, where the majority of epochs 
are clustered over a few years and only three points are found at the beginning of the series. 
While the simple average RV is about 1~\kms\ too large, this is due to the unequal 
distribution of observations, whereas a time-weighted average would support cluster membership. 
I re-classify this star as a binary likely member. 

As I have demonstrated, all four non-members in the \citet{Giampapa2006} sample 
could easily be members, and so I will leave them all in the 
sample.\footnote{While there is good reason for \citet{Geller2015} to employ their more conservative criteria 
in order to produce the cleanest membership list unlikely polluted with interlopers, 
for this work, it is important to check every star we can for signs of excess or deficient 
activity. Incorporating these stars into this analysis does not alter the conclusions 
one would reach if they were rejected.}
Adopting my liberal membership and binarity classifications,
I re-plot the HK index distribution for single members in the 
top-right panel of Figure~\ref{f:hk}
and find that most of the active stars have been removed from the sample. 
This decreases the high-activity frequency from 22\% to 8\% , 
while dropping the sample size from 77 to 51 stars.

Next, I apply the solar color criteria from  \citet{Giampapa2006} in the bottom-left panel to isolate 
the single solar analogs: $0.58 \leq (B-V)_0 \leq 0.76$. 
The sample shrank to 40 stars, but the frequencies basically remained unchanged at 
18\% inactive and 8\% active.

Finally, I apply the 15~\ma\ ISM correction to the \citet{Giampapa2006} 
HK indices for single G dwarfs and re-plot the distribution in the bottom-right panel of Figure~\ref{f:hk}. 
Only one star is now found below solar minimum and I will discuss it in Section~\ref{s:1473}.
Introducing the correction shifts the lower envelope of the distribution 
up to solar minimum 
and raises the sample mean above the solar mean: 204 compared to 194~\ma . 
If M67 is in fact 500 Myr younger than the Sun, 
and if magnetic activity continuously and monotonically decays
according to \citet{mamajek2008}, 
then the solar analogs of M67 should be slightly more active on average 
and, according to this analysis, they are.\footnote{Unfortunately, 
we do not yet have an activity--age relation 
that has been adequately calibrated for middle-aged stars: 
(1) the Sun is the fundamental data point at old age and the calibration of the solar data 
is still being debated and worked out 
(whereas \citeauthor{mamajek2008}~\citeyear{mamajek2008} adopted 
$\overline{S_\odot} = 0.176$, \citeauthor{Egeland2016}~\citeyear{Egeland2016} recently re-calibrated the solar data 
and derived 0.1694 for the average over Cycles 15 to 24); 
(2) the M67 data point was measured from ISM-contaminated spectra; 
(3) there are not yet any other accurate measurements with precise ages older than the Hyades, 
meaning the activity--age relation is essentially an interpolation across 4~Gyr of 
stellar, angular momentum, and dynamo evolution.
Still, if we employ the \citet{mamajek2008} relationship under the assumption that the 
\citet{skumanich1972} relation appears to operate until at least the age of the Sun 
\citep{Meibom2015, Barnes2016, vanSaders2016}, then we 
expect the Sun's activity at 3.5, 4.0, and 4.5~Gyr to be 
$\smw = 0.190, 0.182, 0.175$, and $\lrphk = -4.842, -4.876, -4.908$~dex, respectively. 
These differences are small, but discernible. 
More work is needed to accurately measure the activity of M67's solar analogs and place 
them in the context of somewhat younger clusters (e.g., NGC~752 and Ruprecht~147) 
and older field stars.}

With a standard deviation of 15.2~\ma , the HK distribution of M67's single solar analogs 
still appears wide relative to the Sun, 
but the M67 distribution contains observational errors ranging from 
 5.6 to 19.7~\ma , with a mean and standard deviation of 9.5 and 2.9~\ma . 
The standard deviation of the solar HK distribution is 8.7~\ma . 
Adding the observational error from M67 to the intrinsic width of the 
solar distribution in quadrature yields 12.9~\ma , 
which is still tighter than the M67 distribution. 
Explanations include but are not limited to
(1) perhaps the M67 record better reflects the range of activity the Sun 
exhibits over longer timescales;
(2) perhaps M67's slightly younger solar analogs 
exhibit a wider range of activity due to their relative youth;
(3) the rotation axes of M67's stars are randomly distributed 
relative to our viewpoint on Earth. \citet{Knaack2001} 
calculated that the active Sun's S index would appear diminished by 4\% if 
viewed pole-on \citep[or 208 instead of 219~\ma\ in HK; see also Figure 6b in][]{Shapiro2014} 
The effect is small for an individual star, but does contribute 
to broadening the distribution.
Finally, 
(4) differential ISM absorption contributes an unknown amount of scatter 
to the distribution because we are not yet able to correct the bias 
on a star-by-star basis.

Considering the 21 stars with solar color ($0.63 \leq (B - V)_0 \leq 0.67$), 
13 are single and 8 are binaries including two SB2s. 
Of the binaries, two show RV rms $\approx 2$~\kms\ with no orbital solutions 
and one has a period of 2808~days and a semi-amplitude velocity of 4.5~\kms . 
I consider these three systems to be well-separated and likely immune to 
tidal effects that would otherwise influence the activity and rotation of 
tighter binaries (the remainder have orbital periods of less than 90~days). 
These include S951 (206~\ma ), S1246 (203~\ma ), and S1462 (209~\ma ), 
where the values in parentheses are ISM-corrected HK indices, 
all of which are consistent with the Sun's range.
Adopting these stars into the single-star sample gives 16 stars 
that are directly comparable to the Sun. 
All but one exhibit HK values ranging between 184 and 226~\ma , 
which nearly encompasses the full solar range (once again, 179 to 226~\ma ).

S785 has solar color with $(B-V)_0 = 0.65$, 
and is apparently single (RV rms $= 0.65$~\kms\ and is on the single-star CMD sequence). 
However, it has an ISM-corrected HK index of 237~\ma , 
with values ranging from 216 to 256~\ma\ over three years and four measurements.
Why does S785 appear so active relative to the Sun and M67's solar analogs? 
Clearly, we need to revisit S785 and acquire new activity data.

\subsection{S1473 Is Not a Maunder minimum Candidate}\label{s:1473}

Sanders 1473, $(B-V)_0 = 0.73$, 
is the sole star with a corrected HK index below solar minimum: 
HK = 158 \ma , HK$_{\rm corr}$ = 173 \ma . 
According to \citet{mamajek2008}, $\lrphk\ = -5.13$ for this star 
(conveyed to those authors by M. Giampapa). 
While \citeauthor{Giampapa2006} did not list activity measurements for each epoch, 
the reader can see from their Figure~9  
that S1473 was observed twice \citep[page 457 of][]{Giampapa2006}. 
From that activity versus time of observation plot for S1473, 
I estimate by eye HK = 153 m\AA\ in 1997, and 188 m\AA\ in 2002. 
Introducing the ISM correction adjusts these values up to 168 and 203 \ma .
While S1473 appears to have dipped well below the modern solar minimum, 
it did not remain in an extended minimum, and 
reached a value typical of the Sun just five years later.
S1473 is not an MM candidate because it 
did not remain in a flat, inactive state.


In addition to S1473, there are four 
single solar-color stars with relatively low HK emission:
S801 (182: $\sim$200, $\sim$165),
S991 (179: $\sim$180, $\sim$150, $\sim$193, $\sim$190, $\sim$185),
S1106 (168: $\sim$205, $\sim$130, $\sim$170$\times4$), and 
S1477 (181: $\sim$200, $\sim$180, $\sim$182, $\sim$155, $\sim$185), 
where the values in parentheses are uncorrected HK indices and
followed by individual epoch measurements estimated by eye from 
Figure~9 in \citet{Giampapa2006}. 
Even though their average ISM-corrected values are above solar minimum, 
it is worth looking at the time series to 
check for flat activity records.
In each case, there is at least one medium/high-activity epoch (i.e., $>190$~\ma ), 
which challenges their classification as MM candidates.
Still, S1106 is intriguing and will be important to revisit with new \caiihk\ observations.
Aside from one high epoch in 1996 ($\sim$205~\ma ),
followed by a very low epoch in 1998 \citep[$\sim$130~\ma , 
which is the lowest single HK measurement of any star observed by][]{Giampapa2006}, 
it appeared flat for the final four years at $\sim170$~\ma .
Adding the 15~\ma\ ISM correction brings this to 186~\ma , 
only 4~\ma\ higher than solar minimum and consistent within measurement error.
Given that the median lifetime for solar grand minima inferred from 
the radionuclide record is 60 years \citep{Usoskin2013}, 
we might still be able to catch it in an extended minimum if that is 
the case, since it would have only begun in 1998, some 18 years ago. 
Despite this intriguing possibility, it cannot yet be classified as an MM candidate 
given its high HK epoch in 1996---if not for that first epoch, S1106 
would have been an excellent target for follow-up as an MM candidate. 

Upon investigating each low-activity star, 
I conclude that there is no compelling evidence at this time 
for Maunder minimum candidates in M67.

\subsection{Active stars and multiplicity}
I already demonstrated that the majority of overactive stars 
turned out to be binaries according to \citet{Geller2015}.
A close stellar companion can tidally interact with the primary 
star and alter its angular momentum evolution 
\citep[e.g.,][]{Meibom2005, Meibom2006}.
Such stars can remain rapidly rotating and stay active despite their old age.
This effect certainly accounts for some of the apparently overactive stars.

The location of the foreground ISM features near the edge of the HK index bandpass 
raises a second observational bias issue for binaries. 
As the primary completes its orbit, 
the changing relative velocity between the star and the ISM 
will periodically bring the ISM absorption lines further into and 
back out of the HK bandpass, 
artificially diminishing and raising the activity level. 
This should cause the activity levels of binaries to appear more variable 
than single stars of similar spectral type. 
This could explain why certain binaries appeared to be inactive while others 
appeared to be more active, 
even though most of these systems do not have short periods where 
tides are more effective at preventing stellar spin-down.

\begin{figure*}\begin{center}
\plottwo{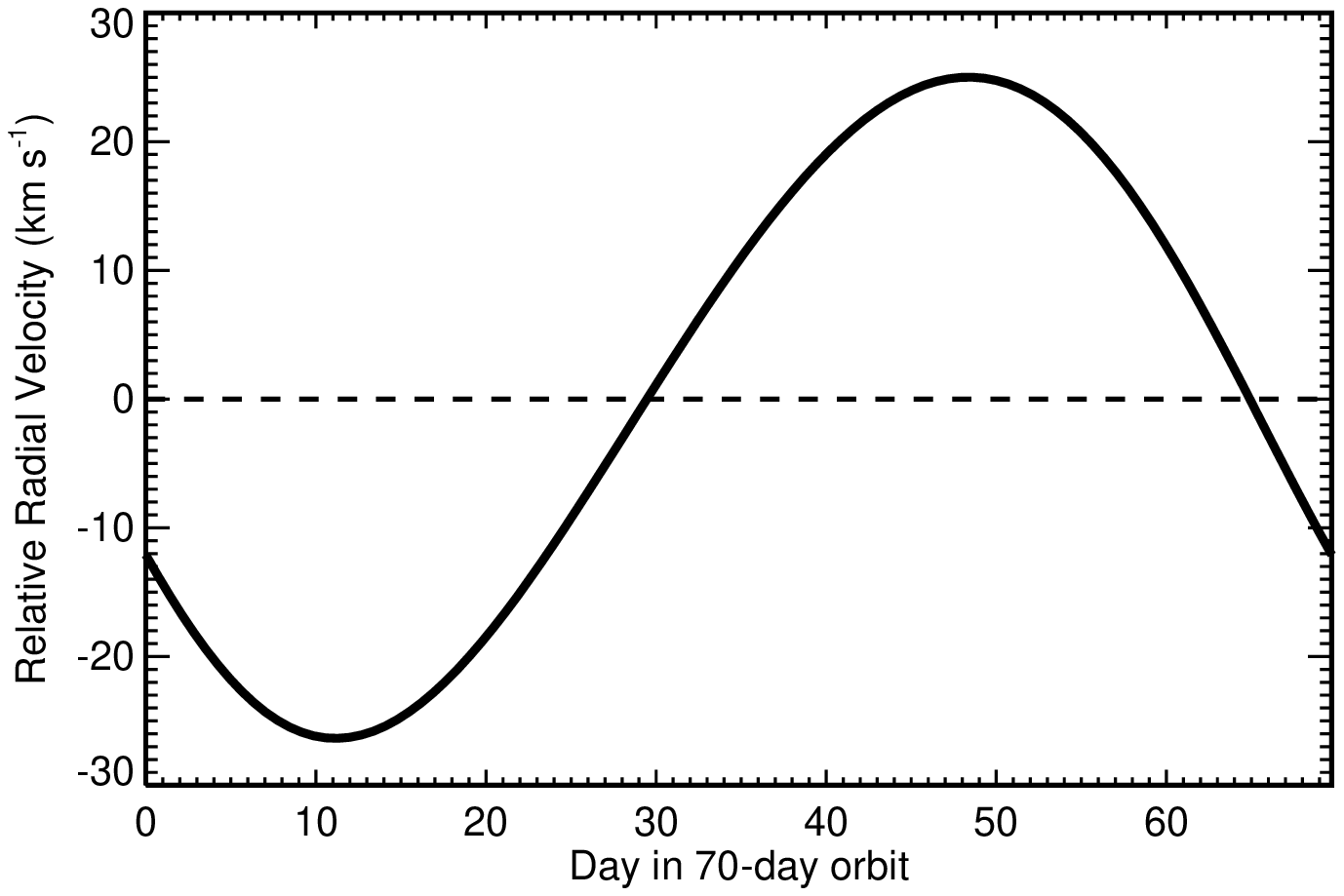}{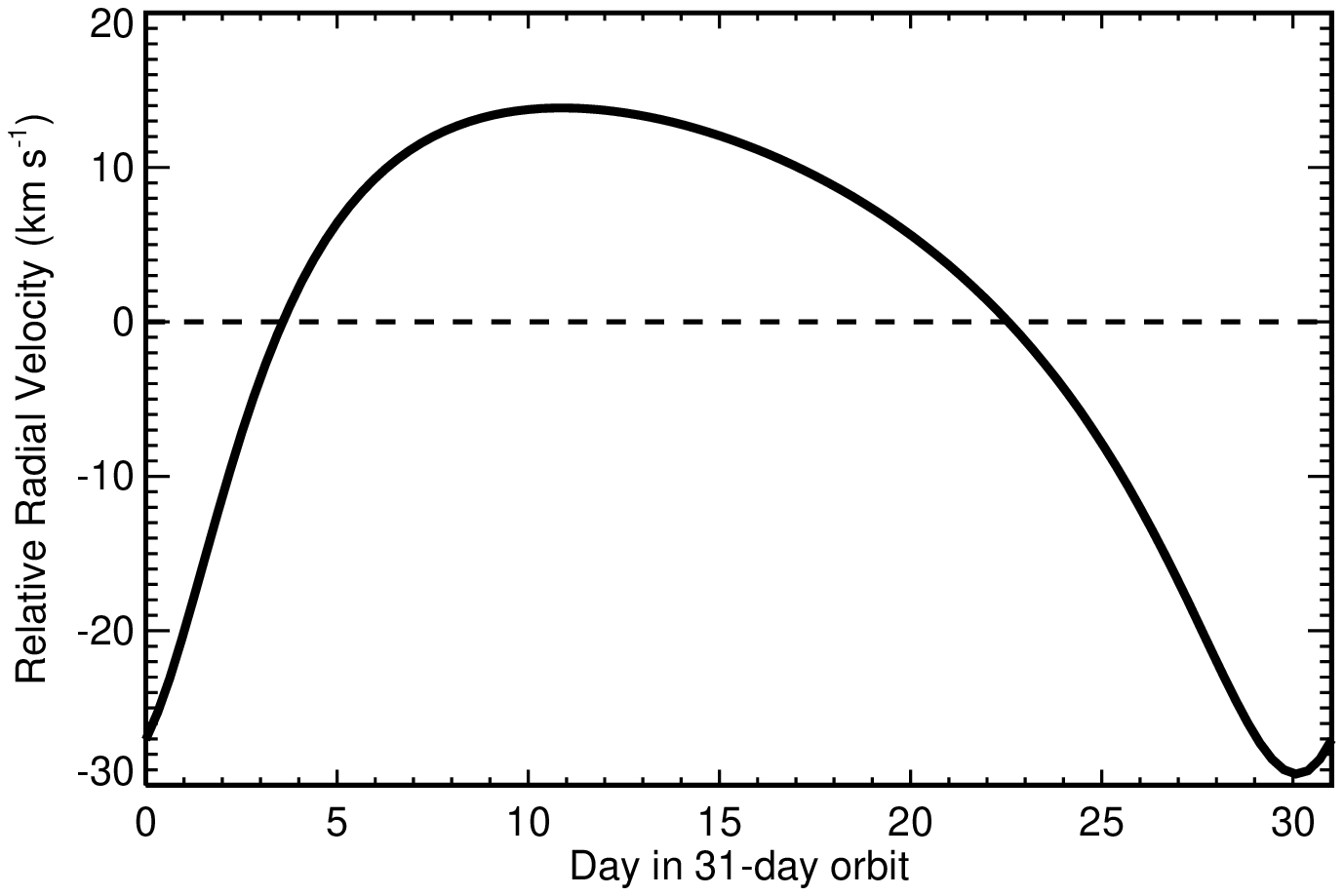}
\plottwo{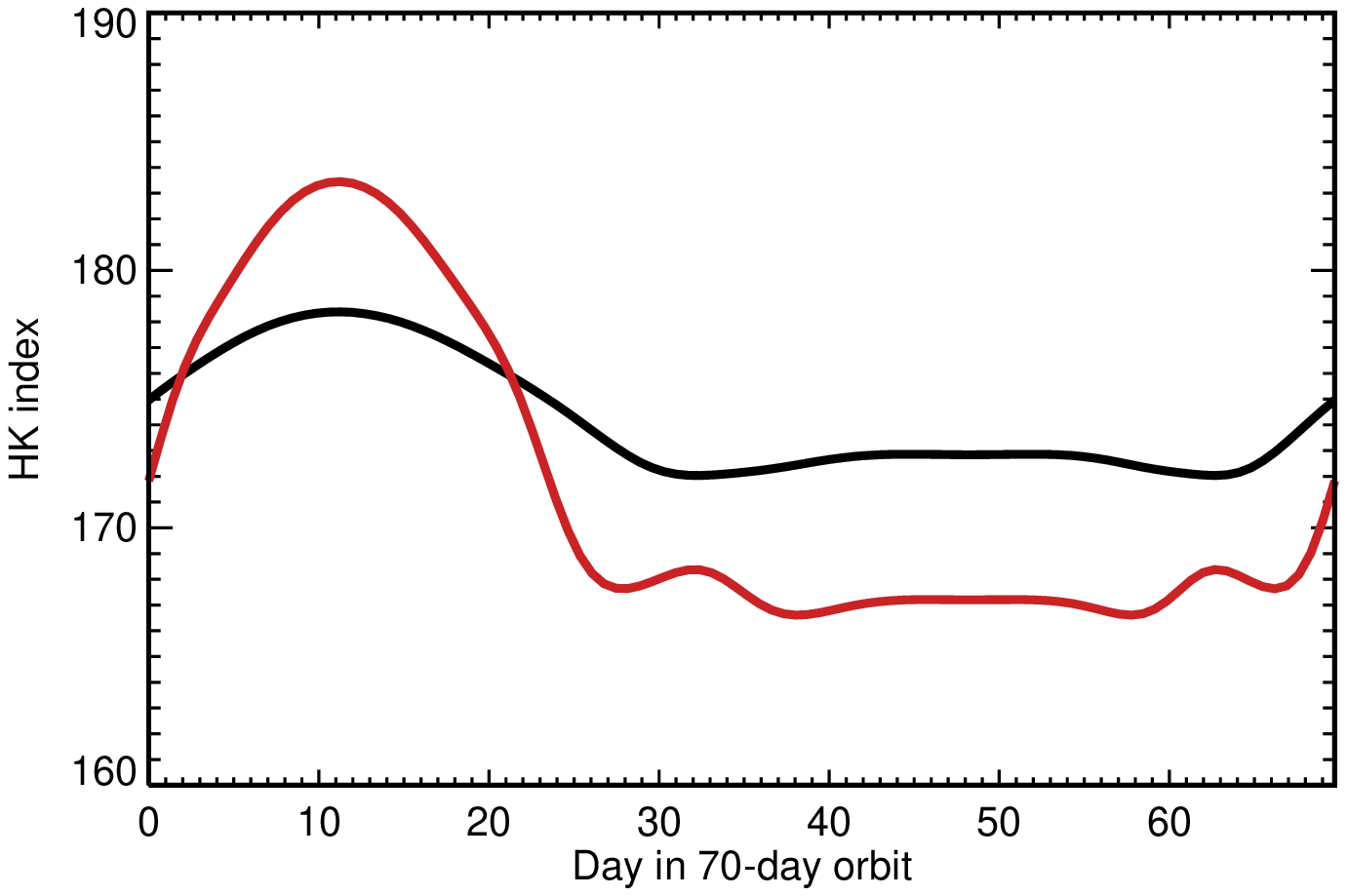}{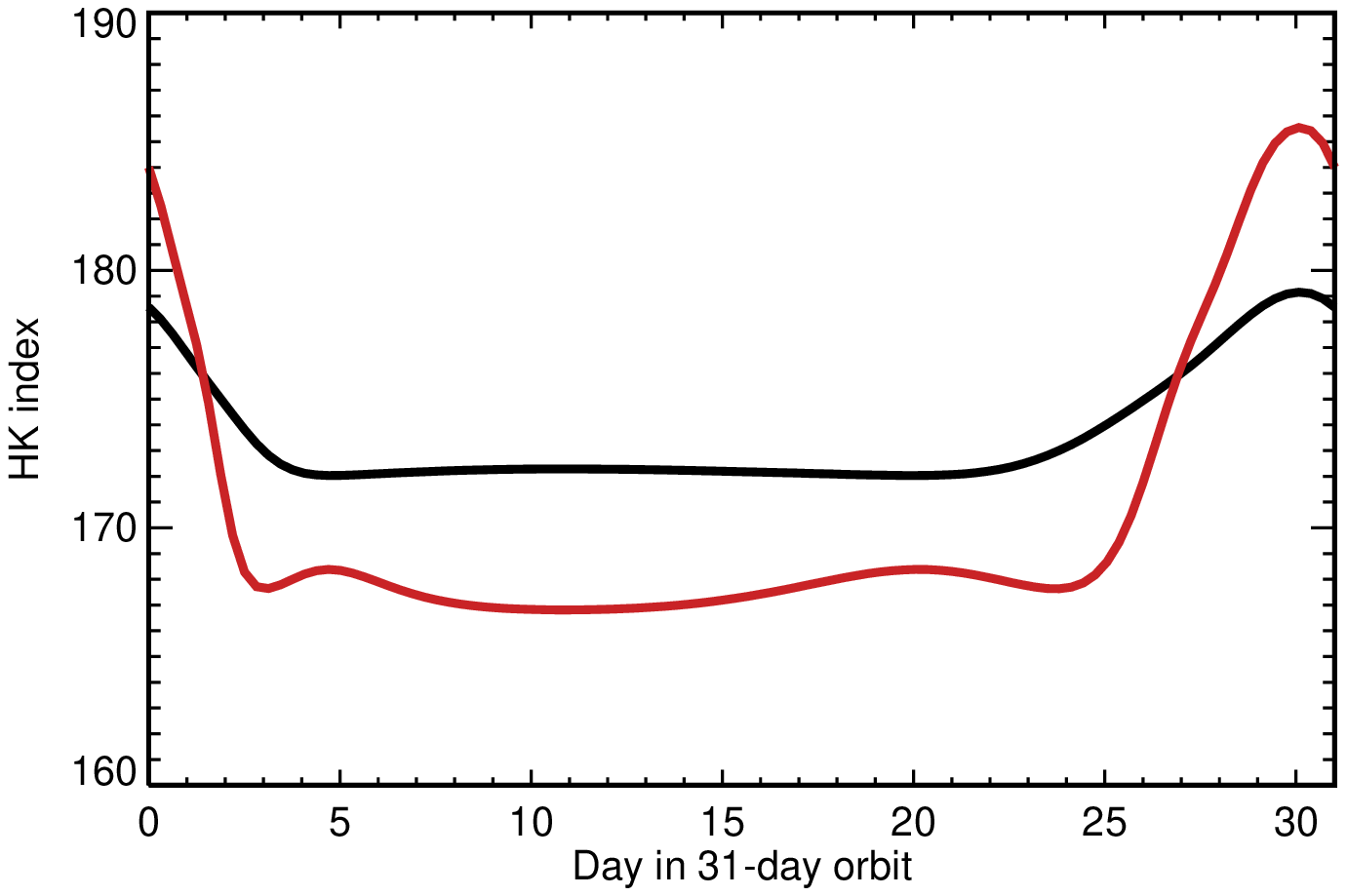}
    \caption{The ISM--SB1 effect: top panels show model RVs of 
    SB1 members S1247 (left) and S943 (right) over 
    the course of their respective orbits; bottom panels show 
    a simulation of the resulting HK index measured from high-resolution (red) 
    and low-resolution spectra (black). 
    A SOLIS ISS spectrum was used as a template, then the M67 ISM model was 
    applied to simulate a M67 solar twin; 
    a Gaussian broadening kernel with FWHM of 0.8~\AA\ was applied to 
    produce the low-resolution version. 
    These figures illustrate how orbital motion can periodically sweep 
    more or less ISM absorption lines into the HK bandpass. 
    Observing at lower spectral resolution tends to mitigate this effect 
    as the spectral resolution (0.8~\AA\ for Hydra) approaches or exceeds the 
    orbital RV motion (which is $\sim 30$~\kms\ in these examples, or $\sim 0.4$~\AA).
    \label{f:SB1}}
\end{center}
\end{figure*}

This issue, which I call the ISM--SB1 effect, is illustrated in Figure~\ref{f:SB1}, 
which plots model RVs of SB1 members S1247 and S943 generated with 
\texttt{RVLIN} \citep{RVLIN} 
from their orbital solutions over the course of their 
respective orbits,\footnote{Orbital solution information provided 
by D.~Latham.}
and shows the corresponding HK indices derived from high-resolution and low-resolution solar spectra from SOLIS/ISS with the M67 ISM applied to simulate a cluster solar twin. 
These panels demonstrate how orbital motion can periodically sweep 
    more or less ISM absorption lines into the HK bandpass. 
    Observing at lower spectral resolution tends to mitigate this effect 
    as the spectral resolution (0.8~\AA\ for Hydra) approaches or exceeds the 
    orbital RV motion (which is $\sim 30$~\kms\ in these examples, or $\sim 0.4$~\AA).

\citet{Giampapa2006} noted that the star S1452, with $(B-V)_0 = 0.62$, 
was apparently single (i.e., no binary detected from RV variation $<$2 \kms ), 
and yet showed an emission core reversal indicative of high activity, 
with $\lrphk\ = -4.35$~dex (this translates to 200~Myr with the 
\citeauthor{mamajek2008}~\citeyear{mamajek2008} activity--age relation).
Following the survey, \citet{ReinersGiampapa2009}
pursued \vsini\ observations with the goal of understanding the 
range of activity levels, and measured $\vsini = 4 \pm 0.5$ \kms for S1452. 
With rotational broadening approximately double that of the Sun, 
the rotation period for S1452 should be at least half of the Sun, 
or $\sim$13 days. 
I downloaded the \textit{K2} light curve for this star 
from the K2SFF dataset on MAST 
\citep[prepared following][]{vj14, vanderburg2016}.\footnote{Available for download at
https://archive.stsci.edu/missions/hlsp/k2sff/html/c05/ep211410278.html}
It shows shows clear spot modulation with an amplitude of $\approx$1.5\% 
and a period of approximately 4.6~days; the discrepancy between \vsini\ and $P_{\rm rot}$ 
is probably due to inclination of the stellar rotation axis.
\citet{ReinersGiampapa2009} attributed this higher activity and rapid rotation
to a reduced braking efficiency for this star. 
At the time of those studies, the star was assumed to be single.
However, \citet{Geller2015} now report that S1425 as a SB1 member. 
Latham found an orbital period $P_{\rm orb} = 358.5$~days, 
with eccentricity $e = 0.196$, and RV semi-amplitude $K = 4.12$~\kms (private communication). 
S1452 appears on the single-star sequence of the M67 $(B-V)$ CMD, 
so the companion must have a relatively low luminosity (e.g., white dwarf or M dwarf). 

S1452 is reminiscent of a particularly overactive solar twin found in the 3~Gyr Ruprecht~147 cluster
that was also observed by \textit{K2}.
Its light curve shows strong spot modulation due to rapid rotation as well as 
transits indicative of a Jupiter-sized object in a short-period orbit. 
Eventually, my collaborators and I characterized the system and identified a warm brown dwarf in an 
eccentric $\sim$5~day orbit as the culprit, 
which we argue is spinning up its host star as its orbit circularizes \citep{Curtis2016}.\footnote{See also 
\citet{Nowak2017} for an independent discovery and characterization of the system.}

So far, no transit has been detected for S1452 from the \textit{K2} light curve.
Additionally, there is no significant RV scatter about the approximately one year orbit indicative 
of a brown dwarf; however, the RV surveys to date might not have achieved sufficient precision 
to tease out a hot Jupiter companion, especially if we happen to enjoy an unfavorable viewing angle.
The nature and history of S1452 and its unknown stellar companion continue to defy explanation.

\section{Summary}

The stars of M67 provide a window into the Sun's behavior on timescales that greatly 
surpass the sunspot and radionuclide records. 
If M67's age is really 500 Myr younger than the Sun's, perhaps the cluster's stars 
actually reveal the Sun's magnetic activity during the Cambrian explosion, 
the period of rapid evolution and proliferation of animal life. 
Given the impact of magnetic activity on biological evolution, habitability, and climate, 
the work of \citet{Giampapa2006} and similar studies are absolutely crucial for understanding 
this important relationship. 
Furthermore, measuring accurate activity levels as a function of stellar mass, both the mean and variation, 
is useful for empirically age-dating isolated Sun-like stars. 
This is especially important for studying exoplanets, their habitability, 
and for predicting their atmospheric biosignatures based on 
Earth's own particular evolution.

\cite{Pace2004} recognized the important contaminating effect of the ISM on chromospheric activity indices. 
I developed their work further by demonstrating the variability of the ISM across M67, 
and by explaining the implications of the ISM bias on results of the ground-breaking survey of \citet{Giampapa2006}. 
Using the updated membership and multiplicity catalog of \citet{Geller2015}, 
I find that the majority of the high-activity stars of M67 are actually binaries. 
By applying an ISM model derived from a cluster blue straggler, I 
calculated an HK index correction, which I added to the M67 measurements. 
All but one star now show an activity level on par with or in excess of the solar minimum. 
I explained that the sole star with a two-epoch activity mean below solar minimum is 
not a Maunder minimum candidate, because its activity level recovered from a low in 1997 to  
a value approximately equal to the solar average by 2002.
Other single stars with average levels near solar minimum also show at least one 
single-epoch HK value in line with the solar average.

Due to the uncertainty surrounding the Sun's behavior during the Maunder Minimum, 
including the value and variability of the total solar irradiance and chromospheric and coronal emission, 
I cannot issue a definitive statement on the actual frequency of MM stars in M67 at this time. 
However, \citet{Giampapa2006} originally suggested that that these stars might be in low-activity states analogous to the MM because of the presence of the sub-solar activity tail of M67's HK distribution. Given that this is likely not true in light of the ISM correction, there is no longer reason to suggest that these stars are in MM states. The fact that the low-activity stars in the M67 distribution also appear variable in the \citet{Giampapa2006} time series also contradict their classification as MM candidates 
according to the criteria I adopted in this study.
Therefore, I concluded that there is no evidence for stars in Maunder minimum states in M67 at this time. 


Applying the ISM correction also shifted the cluster's average activity upward so that 
one-third of the sample now exceeds solar maximum.
This means that the Sun was possibly more active on average during the Cambrian explosion, 
though this conclusion requires a more accurate and precise age for M67 relative to the Sun.
A recent asteroseismic analysis of the cluster's red giants observed with \textit{K2} 
determined the age of M67 to be $3.46 \pm 0.13$~Gyr, which is even younger than the canonical 4~Gyr value
\citep{Stello2016}.

In the future, it will be important to map the interstellar \caii\ lines across the cluster so that 
activity levels for individual stars can be more accurately corrected. 
I intend to pursue this using the full sample of M67 spectra from VLT UVES, 
along with TRES spectra of the cluster's 13 blue stragglers \citep{Latham1996}.
These ISM models should be applied to the \caiihk\ spectra directly, 
to properly account for stellar binarity, spectral type, and activity level.

Acquiring high-resolution spectra at high $S/N$ would provide much needed clarity and 
would demonstrate whether or not M67's solar analogs truly are more active than the Sun is, on average, 
because we would be able to directly compare solar and M67 emission profiles without 
needing to worry about calibration differences or if our extracted indices are on a consistent system.
In particular, I recommend observing the following stars showing 
low activity (S1473, S1106), 
high activity (S785)
and extreme activity (S1452), 
in addition to the ``well-behaved'' solar twins.
M67 constitutes a fundamental pillar for stellar and solar astrophysics. 
As a community, we should value activity surveys of this cluster and ensure 
that new epochs continue to be gathered.

\acknowledgments

I performed the majority of this work while serving as a Predoctoral Fellow of the 
Smithsonian Astrophysical Observatory at the Harvard--Smithsonian Center for Astrophysics 
during the 2015--2016 academic year, which was my final year enrolled in the 
Ph.D. program at the Pennsylvania State University.
I responded to my referee's report and finalized this manuscript 
after starting as an Astronomy \& Astrophysics 
Postdoctoral Fellow of the National Science Foundation at Columbia University.

I would like to thank Dave Latham for 
providing the TRES spectrum of F280 and 
sharing his RV and orbital solution information for the M67 binaries. 
I thank Andrew Vanderburg for sharing his \textit{K2} light curve for S1452 with me before 
it was released to the public.
I am also grateful to Mark Giampapa for enlightening conversations about magnetic activity 
and its observation in old clusters, 
for his continued dedication to understanding the evolution of the Sun and Sun-like stars, 
and for working with me to pursue chromospheric activity 
observations of the 3~Gyr cluster Ruprecht 147 with WIYN Hydra from the NASA--NSF EXPLORE program allocated by NOAO.
I appreciate comments on my draft manuscript from 
my Ph.D. adviser, Jason T. Wright; members of my 
Penn State dissertation committee, including 
Ron Gilliland, Robin Ciardullo, Richard Wade, Rebekah Dawson, and G. Jogesh Babu; 
and members of my SAO predoctoral committee including \soren\ and Steve Saar. 
Finally, I would like to thank the referee for their detailed comments, 
which helped me understand the broader context of the issues I am beginning to address in my research
as well as to improve this manuscript.

My work is supported by the National Science Foundation 
Astronomy and Astrophysics Postdoctoral Fellowship under award AST-1602662, 
the National Aeronautics and Space Administration under
grant NNX16AE64G issued through the \textit{K2} Guest Observer Program. 
and the NASA--NSF Partnership for Exoplanet Observational Research (NN-EXPLORE) 
awarded to NOAO program 2016A-0298.

The Center for Exoplanets and Habitable Worlds is supported by the
Pennsylvania State University, the Eberly College of Science, and the
Pennsylvania Space Grant Consortium.

This work utilized observations made with ESO Telescopes at 
the La Silla Paranal Observatory under program IDs 
065.L-0427, 066.D-0457, 068.D-0332, 068.D-0491, 069.D-0454, 070.D-0421, and 079.C-0131, 
and acquired from the ESO Science Archive Facility. 

This work also utilized SOLIS data obtained by the NSO Integrated Synoptic Program (NISP), 
managed by the National Solar Observatory, 
which is operated by the Association of Universities for Research in Astronomy (AURA), Inc., 
under a cooperative agreement with the National Science Foundation.

This research made use of NASA's Astrophysics Data System and
and the VizieR and SIMBAD databases operated at CDS, Strasbourg, France.

\vspace{5mm}
\facilities{SOLIS (ISS), VLT:Kueyen (UVES), FLWO:1.5m (TRES), \textit{Kepler} (\textit{K2})}

\end{document}